\documentclass[ALICE,manyauthors]{cernphprep}

\usepackage[comma,square,numbers,sort&compress]{natbib}
\usepackage{hyperref}
\usepackage{lineno}
\usepackage{subfigure}
\usepackage{color}

\newcommand{\acceff}{\ensuremath{A \cdot \epsilon}\xspace}
\newcommand{\ci}[1]{Ref.~\cite{#1}\xspace}
\newcommand{\centr}[2]{\mbox{#1--#2\%}\xspace}

\newcommand{\dimuon}{\ensuremath{\mu^+ \, \mu^-}\xspace}
\newcommand{\dimuonLS}{\ensuremath{\mu^{\pm} \, \mu^{\pm}}\xspace}

\newcommand{\etamu}{\ensuremath{\eta}\xspace}
\newcommand{\fig}[1]{\figurename~\ref{#1}\xspace}
\newcommand{\fmc}{\mbox{fm/}\ensuremath{c}\xspace}
\newcommand{\fnorm}[1][]{\ensuremath{F_{#1\mu\text{-trig/MB}}}\xspace}

\newcommand{\GeVc}{~\mbox{GeV/}\ensuremath{c}\xspace}
\newcommand{\GeVcc}{~\mbox{GeV/}\ensuremath{c^2}\xspace}

\newcommand{\invmass}{\ensuremath{M_{\mu\mu}}\xspace}

\newcommand{\lumi}[1][n]{~\ensuremath{\text{#1b}^{-1}}\xspace}

\newcommand{\ncoll}{\ensuremath{N_\text{coll}}\xspace}
\newcommand{\npart}{\ensuremath{N_\text{part}}\xspace}
\newcommand{\npartNcoll}{\ensuremath{\langle N_\text{part} \rangle_{N_\text{coll}}}\xspace}

\newcommand{\PbPb}{Pb-Pb\xspace}

\newcommand{\pPb}{p-Pb\xspace}
\newcommand{\pt}{\ensuremath{p_{\textrm{T}}}\xspace}
\newcommand{\ptmu}{\ensuremath{p_{\textrm{T}}}\xspace}
\newcommand{\qqbar}[1]{\mbox{#1}\ensuremath{\overline{\mbox{#1}}}\xspace}
\newcommand{\Raa}{\ensuremath{R_\text{AA}}\xspace}
\newcommand{\snn}{\ensuremath{\sqrt{s_{\mathrm{NN}}}}\xspace}
\newcommand{\sect}[1]{Sect.~\ref{#1}\xspace}
\newcommand{\taa}{\ensuremath{\langle T_{\rm AA} \rangle}\xspace}
\newcommand{\tab}[1]{\tablename~\ref{#1}\xspace}

\newcommand{\wBoson}{\ensuremath{\text{W}^{\pm}}\xspace}
\newcommand{\zBoson}{\ensuremath{\text{Z}^{0}}\xspace}

\graphicspath{{plots/}}

\begin{document}%

\begin{titlepage}
\PHyear{2017}
\PHnumber{305}      
\PHdate{28 November}  
%

\title{Measurement of \zBoson-boson production at large rapidities in \PbPb~collisions at ${\mathbf {\snn=5.02}}$~TeV}
\ShortTitle{\zBoson-boson production in \PbPb~collisions at $\snn=5.02$~TeV}   

\Collaboration{ALICE Collaboration\thanks{See Appendix~\ref{app:collab} for the list of collaboration members}}
\ShortAuthor{ALICE Collaboration} 

\begin{abstract}

The production of \zBoson bosons at large rapidities in \PbPb~collisions at $\snn=5.02$~TeV is reported. 
\zBoson candidates are reconstructed in the dimuon decay channel ($\zBoson \rightarrow \dimuon$), based on muons selected with pseudo-rapidity $-4.0<\etamu<-2.5$ and $\ptmu>20$\GeVc. 
The invariant yield and the nuclear modification factor, \Raa, are presented as a function of rapidity and collision centrality.
The value of \Raa for the \centr{0}{20} central \PbPb collisions is $0.67 \pm 0.11 \, \mbox{(stat.)} \, \pm 0.03 \, \mbox{(syst.)} \, \pm 0.06 \, \mbox{(corr. syst.)}$, exhibiting a deviation of $2.6 \sigma$ from unity.
The results are well-described by calculations that include nuclear modifications of the parton distribution functions, while the predictions using vacuum PDFs deviate from data by $2.3\sigma$ in the \centr{0}{90} centrality class and by $3\sigma$ in the \centr{0}{20} central collisions.

\end{abstract}
\end{titlepage}
\setcounter{page}{2}


\section{Introduction}

\zBoson bosons are weakly interacting probes formed early in the evolution of hadronic collisions ($t_f \sim 1/M \ll 0.01$~\fmc), with a typical decay time $t_d \sim 0.1$~\fmc.
Their leptonic decays are of particular interest in heavy-ion collisions, since leptons do not interact strongly and their in-medium energy loss by bremsstrahlung is negligible~\cite{Peigne:2007sd}.
\zBoson-boson production rates in hadronic collisions are well-understood, and their measurement via leptonic decays therefore serves as a valuable medium-blind reference for hard processes in heavy-ion collisions~\cite{Kartvelishvili:1995fr,ConesadelValle:2009vp}.

%
%
\zBoson-boson properties have been extensively studied at LEP (CERN), SLC (SLAC), Tevatron (FNAL) and LHC (CERN) in $e^+ e^-$, \qqbar{p} and pp collisions~\cite{
Albajar:1987yz, Alitti:1991dm, 
Abe:1995bm, Abbott:1999tt, Abulencia:2005ix, Aad:2010yt, Aad:2011dm, Aad:2016naf, 
CMS:2011aa, Chatrchyan:2014mua, Aaij:2015gna, Aaij:2015zlq}.
\zBoson-boson production in hadronic collisions is well-described by perturbative Quantum Chromodynamics (pQCD) calculations at next-to-next-to-leading order (NNLO)~\cite{Martin:1999ww,Baur:1998kt}, and their comparison with data provides constraints on Parton Distribution Functions (PDFs)~\cite{Rojo:2015acz,Butterworth:2015oua}. 
%
%
In heavy-ion collisions, \zBoson-boson production can be affected by initial-state effects. 
As a result of the different balance of the number of u and d valence quarks in protons and in lead nuclei (isospin), the yield of \zBoson bosons in \PbPb~collisions at $\snn=5.02$~TeV is expected to increase relative to that in pp collisions by $5-8\%$ at large rapidities, and decrease by $3\%$ at central rapidities~\cite{Paukkunen:2010qg}. 
Modifications of the PDFs in nuclei (nPDFs)~\cite{Kusina:2016fxy, Eskola:2016oht, deFlorian:2003qf, Hirai:2007sx, AtashbarTehrani:2012xh, Khanpour:2016pph, Vogt:2000hp}
 introduce a rapidity-dependent change in yield, with a decrease in yield relative to that in pp collisions of $8-15\%$ at large rapidities, corresponding to the Bjorken-$x$ ranges $x_1\gtrsim 10^{-1}$ and $x_2 \lesssim 10^{-3}$, and an increase by $3\%$ at central rapidity, corresponding to $x_{1,2}\sim 10^{-2}$~\cite{Paukkunen:2010qg,Kusina:2016fxy}. 
The yield could also depend upon effects such as multiple scattering and medium-induced bremsstrahlung of the initial partons in large nuclei~\cite{Neufeld:2010dz}. 

%
The ATLAS, ALICE, CMS and LHCb collaborations have reported measurements of \wBoson- and \zBoson-boson production in \pPb collisions at $\snn=5.02$~TeV~\cite{
Aad:2015gta, Khachatryan:2015pzs, 
Khachatryan:2015hha, 
Alice:2016wka, 
Aaij:2014pvu}, with complementary rapidity coverage.
These measurements are well-described by next-to-leading order (NLO) pQCD calculations~\cite{Paukkunen:2010qg} and by NNLO calculations using the Fully Exclusive W and Z Production code (FEWZ)~\cite{Gavin:2010az}, utilising both nPDFs~\cite{Alice:2016wka} and vacuum PDFs.
The forward-backward asymmetry of \wBoson-boson production suggests the presence of nuclear modification of PDFs~\cite{Khachatryan:2015hha}. 
This sensitivity to nuclear effects indicates the need to include these data in the future nPDF fits. 
%

In \PbPb collisions, \wBoson- and \zBoson-boson measurements at $\snn=2.76$~TeV have been carried out at central rapidity by the ATLAS and CMS collaborations~\cite{Aad:2014bha, Aad:2012ew, Chatrchyan:2014csa, Chatrchyan:2012nt}. 
Preliminary \zBoson-boson measurements in \PbPb collisions at $\snn=5.02$~TeV at central rapidity have also been reported recently by ATLAS~\cite{ATLAS:2017zkv}.
The \wBoson- and \zBoson-boson nuclear modification factor, \Raa, defined as the ratio of the yields in \PbPb collisions and the cross-section in pp collisions normalised by the nuclear overlap function \taa, which represents the effective overlap area of the two interacting nuclei~\cite{Miller:2007ri}, is measured to be consistent with unity within uncertainties, with no centrality dependence~\cite{Chatrchyan:2014csa,Chatrchyan:2012nt,ATLAS:2017zkv}.

Measurements at high collision energy and large rapidities are sensitive to low Bjorken-$x$ processes, and are therefore important to further constrain the initial-state effects on electroweak boson production and to establish a reference for medium-sensitive observables. 

This paper presents the first measurement of \zBoson-boson production in \PbPb~collisions at $\snn=5.02$~TeV at large rapidities. 
Opposite-sign muon pairs from \zBoson-boson decays with $2.5<y<4.0$\footnote{In the ALICE reference frame the muon spectrometer covers a negative $\eta$ range and, consequently, a negative $y$ range. However, since the \PbPb system is symmetric in rapidity, a positive $y$ notation is used to present the results.} are measured with the ALICE detector. 
The yield of \dimuon pairs includes contributions from virtual-photon processes and from their interference effects.
This measurement probes the nPDFs of large-$x$ valence quarks ($x_1\gtrsim 10^{-1}$) and low-$x$ sea quarks ($x_2 \lesssim 10^{-3}$) at $Q^2 \sim M_{\rm Z}^2$.
The invariant yields and \Raa are reported as a function of rapidity and collision centrality. 
The results are compared to model calculations including nPDFs.
These measurements complement the measurements in \pPb~collisions at $\snn=5.02$~TeV at large rapidities~\cite{Alice:2016wka,Aaij:2014pvu}, 
providing increased precision and new information on rapidity and centrality dependence. 
The combination of these results with future \wBoson measurements in a similar kinematic interval will provide constraints on the flavor dependence of nPDFs, in particular the strange quark contribution~\cite{Kusina:2016fxy}. 

This letter is organised as follows: 
the experimental setup and data sample are described in \sect{sec:det_data}; the analysis procedure is presented in \sect{sec:analysis}; the results are presented in \sect{sec:results}; and a summary is given in \sect{sec:conclusion}. 


\section{Experimental setup and dataset}
\label{sec:det_data}

The ALICE detector is described in detail in \ci{Aamodt:2008zz}.
\zBoson bosons are reconstructed via their muonic decay with the ALICE muon spectrometer, which provides muon trigger, tracking and identification in the pseudo-rapidity range $-4.0<\etamu<-2.5$.
The muon spectrometer, as seen from the interaction point, consists of a front absorber of 10 interaction lengths ($\lambda_{\rm int}$) thickness, which reduces the contamination of hadrons and muons from the decay of light particles; five tracking stations; an iron absorber with thickness 7.2~$\lambda_{\rm int}$; and two trigger stations.
Each tracking  station is composed of two planes of multi-wire proportional chambers with cathode-plane readout, while each trigger station consists of two planes of resistive plate chambers.
The third tracking station is located inside the gap of a dipole magnet, which provides a 3~T$\cdot$m magnetic field integral.
The muon spectrometer is completed by a beam shield surrounding the beam pipe that protects the apparatus from secondary particles produced in the interaction of large-$\eta$ primary particles with the pipe itself.

The interaction vertex is reconstructed using the two cylindrical layers of the Silicon Pixel Detector, located at a radial distance of 3.9 and 7.6~cm from the beam axis and covering $|\eta|<2$ and $|\eta|<1.4$, respectively. 
The V0 detector, consisting of two arrays of scintillator counters covering $2.8<\eta<5.1$ and $-3.7<\eta<-1.8$, is used for triggering and evaluation of collision centrality.
Finally, the Zero Degree Calorimeter, placed at 112.5~m from the interaction point along the beam line, is used to reject electromagnetic interactions~\cite{ALICE:2012aa}.

The dataset used in this analysis consists of \PbPb~events at $\snn=5.02$~TeV selected with a dimuon trigger that requires the coincidence of a minimum-bias (MB) trigger and a pair of tracks with opposite sign in the muon spectrometer, each with $\pt \gtrsim 1$\GeVc.
The MB trigger is defined by the coincidence of the signals from both arrays of the V0.
The MB trigger is fully efficient for events within the \centr{0}{90} centrality interval, which are used in this analysis. 
The muon trigger efficiency has a plateau of about 98\% for muons with $\ptmu>5$\GeVc. 
The resulting efficiency for pairs of opposite-sign muons, with muon $\ptmu>20$~\GeVc and $-4.0<\etamu<-2.5$, is 95\%.
After all event selection cuts, the dataset corresponds to an integrated luminosity of about 225\lumi[$\mu$].

\section{Analysis procedure}
\label{sec:analysis}

The procedure for \zBoson-boson signal extraction in this analysis is the same as that used in the analysis of \pPb collisions at $\snn=5.02$~TeV~\cite{Alice:2016wka}.
Tracks are reconstructed in the muon spectrometer using the algorithm described in \ci{Aamodt:2011gj}.
Tracks are selected for analysis if they have pseudorapidity $-4.0<\eta<-2.5$ and polar angle $170^\circ<\theta_{\rm abs}<178^\circ$, measured at the end of the front absorber. 
This selection rejects particles that cross the high-density region of the front absorber and undergo significant multiple scattering.
Tracks reconstructed in the tracking stations are identified as muons if they match a track segment in the trigger stations, placed downstream the iron wall.
The contamination from background tracks that do not point to the interaction vertex is reduced by utilising the product of the momentum and the distance of closest approach to the interaction vertex.
This cut removes 88\% of all tracks for events in the \centr{0}{90} centrality interval, while retaining all signal candidates with negligible residual background contribution. 

\begin{figure}[ht]
\centering

\subfigure[]{
	\label{fig:ZRawYield_0_90}
	\includegraphics[width=0.48\textwidth]{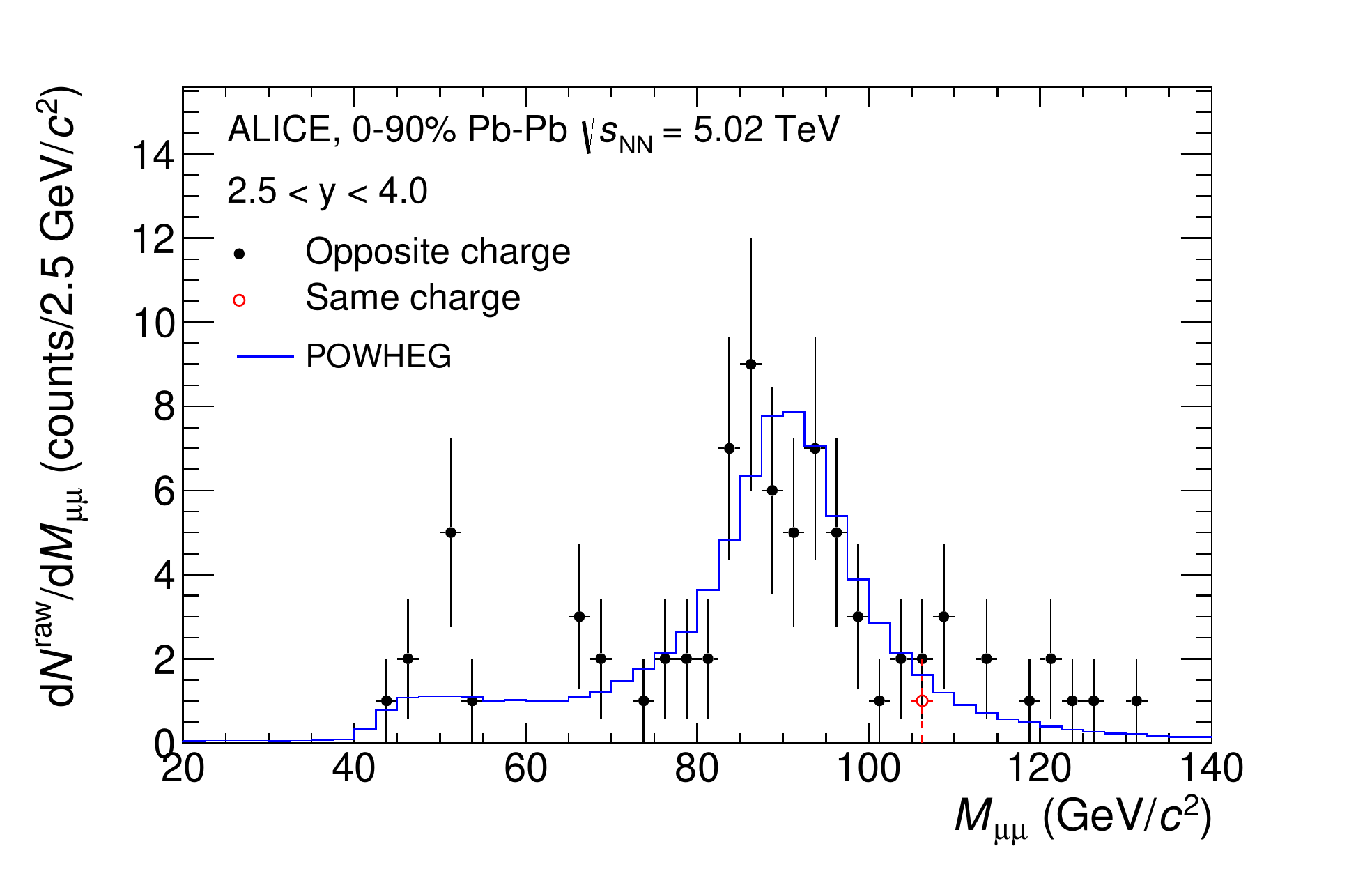}
	} 

\subfigure[]{
	\label{fig:ZRawYield_0_20}
	\includegraphics[width=0.48\textwidth]{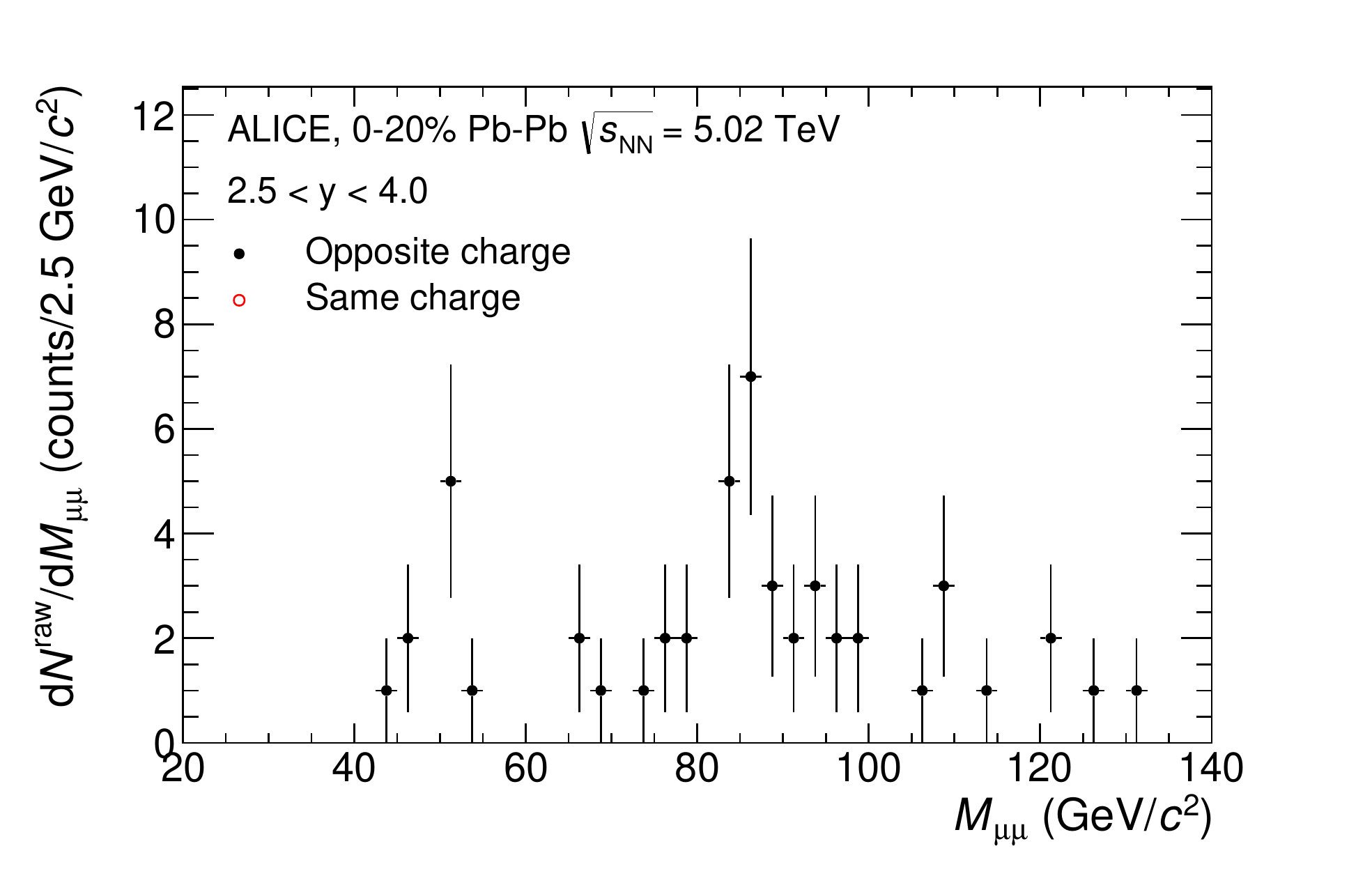}
	} 
\subfigure[]{
	\label{fig:ZRawYield_20_90}
	\includegraphics[width=0.48\textwidth]{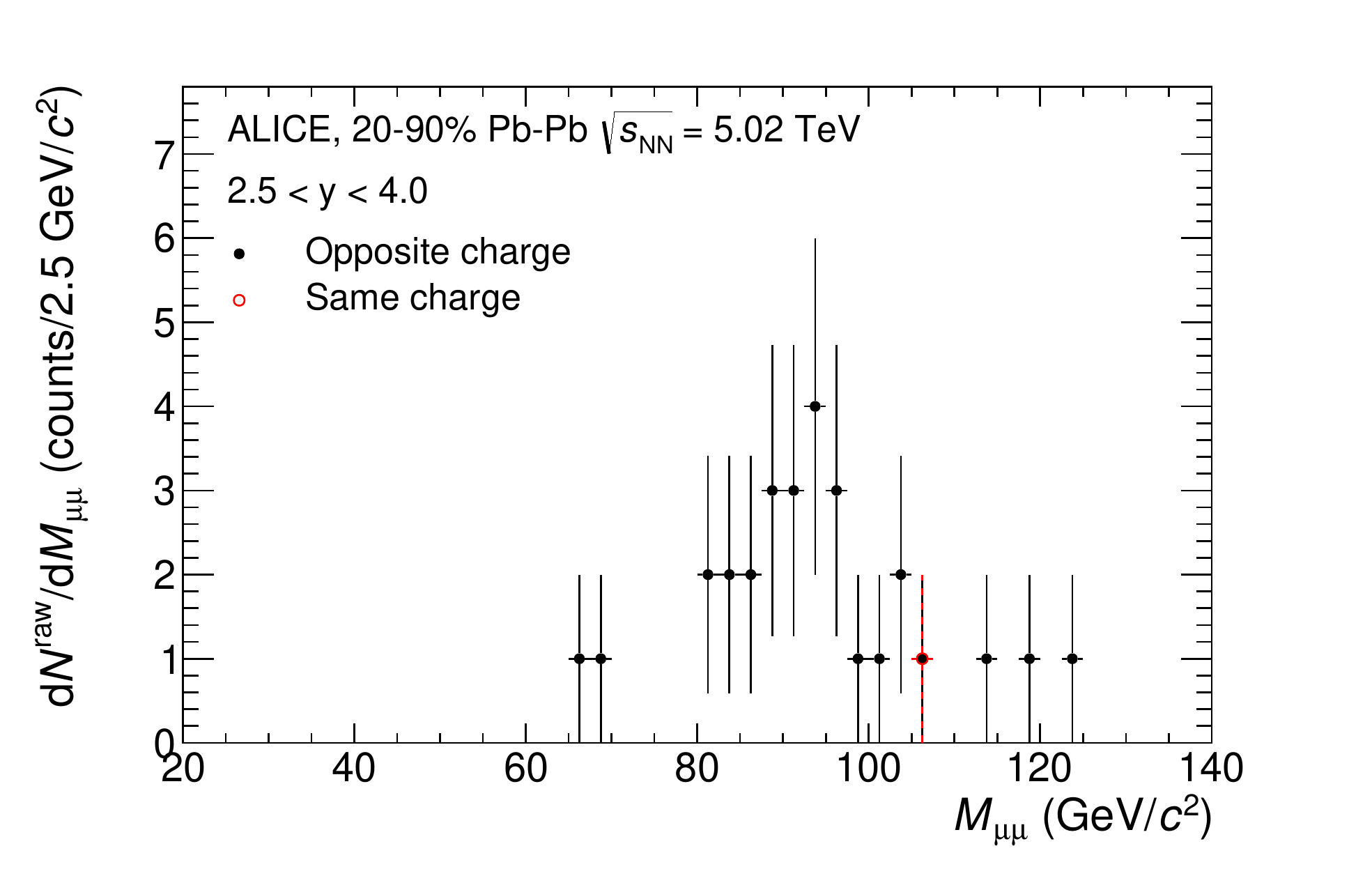}
	} 
\caption{
Invariant mass distribution of \dimuon pairs for \PbPb collisions at $\snn=5.02$~TeV, reconstructed using muons with $-4.0<\etamu<-2.5$ and $\ptmu>20$\GeVc (black points).
The panels present the distribution in different centrality intervals. The error bars are of statistical origin only.
The invariant mass distribution of like-sign muon pairs is also shown (red open points). Only one like-sign candidate is found in the \centr{20}{90} centrality interval.
The solid blue line drawn in \fig{fig:ZRawYield_0_90} represents the distribution from a POWHEG simulation for \PbPb collisions without nuclear modification of PDFs (see text for details).
}
\label{fig:ZRawYield}
\end{figure}

Only muons with $\pt > 20\GeVc$ are used in this analysis.
This selection reduces the contribution of muons from the decay of charm, beauty and low-mass resonances (see below).
\zBoson-boson candidates are formed by combining pairs of opposite-sign muons.
The candidates are further selected by requiring that their rapidity, calculated using the measured invariant mass, is in the interval $2.5<y<4.0$.
Figure~\ref{fig:ZRawYield} presents the \dimuon invariant mass distribution in the centrality intervals \centr{0}{90} in \fig{fig:ZRawYield_0_90}, \centr{0}{20} in \fig{fig:ZRawYield_0_20}, and \centr{20}{90} in \fig{fig:ZRawYield_20_90}.
The distribution for the \centr{0}{90} centrality interval is compared with the result of a Monte Carlo (MC) simulation obtained using the POWHEG~\cite{Alioli:2008gx} event generator paired with PYTHIA~6.4.25~\cite{Sjostrand:2006za} for the parton shower.
The propagation of the particles through the detector is simulated with the GEANT3 code~\cite{Brun:1994aa}.
The isospin of the Pb-nucleus is accounted for by a weighted average of neutron and proton interactions, but no modification of the nucleon PDF was applied to account for nuclear effects.
The simulations account for variations in the detector response with time and in-situ alignment effects. A data-driven description of the muon momentum resolution is also implemented (see \ci{Alice:2016wka} for details). 
The shape of the \dimuon invariant mass distribution, which is mainly affected by the momentum resolution, is similar in data and MC. 

%
%
Various background sources contribute to the \dimuon invariant mass distribution.
Contamination from the decay of \qqbar{t} ($\qqbar{t} \longrightarrow \dimuon \, X$) and $\tau$ ($\zBoson \longrightarrow \tau \tau \longrightarrow \dimuon \, X$) pairs is estimated with POWHEG simulations~\cite{Alioli:2008gx,Aad:2011dm,Chatrchyan:2011wt} and found to be smaller than 0.5\% of the signal yield, which is considered as a systematic uncertainty.
The contribution of opposite-sign muon pairs from the decay of \qqbar{c} ($\qqbar{c} \longrightarrow \dimuon \, X$)  and \qqbar{b} ($\qqbar{b} \longrightarrow \dimuon \, X$)  pairs was studied in \pPb collisions~\cite{Alice:2016wka} and found to be smaller than that of  \qqbar{t} and $\tau$ pairs.
In \PbPb collisions, the presence of high-\pt muons from the decay of heavy-flavour pairs is expected to be further reduced due to the in-medium energy loss of heavy quarks.
This contribution was therefore neglected.
Finally, the combinatorial contribution from the random pairing of muons in the event is evaluated via like-sign muon pairs (\dimuonLS).
This combinatorial contribution is found to be small (one candidate in the \centr{20}{90} centrality interval) and is subtracted from the signal estimate.

%
%
The number of \zBoson candidates is estimated using the procedure described in Ref.~\cite{Alice:2016wka}, by counting the entries in the \dimuon invariant mass interval $60<\invmass<120$\GeVcc after subtracting the contribution from like-sign pairs for each centrality and rapidity interval.
A total of 64 candidates is found in the \centr{0}{90} centrality bin, of which 37 are in the \centr{0}{20} bin and 27 in the \centr{20}{90} bin. 
As a function of rapidity, 33 candidates are in the interval $2.5<y<3.0$, and 31 are in the interval $3.0<y<4.0$.
The raw yields are corrected for the detector acceptance and for reconstruction and selection efficiency (\acceff).
The value of \acceff is 75\% for events in the \centr{0}{90} centrality interval, estimated using the POWHEG~\cite{Alioli:2008gx} simulations described previously.
The dependence of the efficiency on the detector occupancy was evaluated by embedding the generated \zBoson signal in real MB \PbPb data.
The \acceff term is constant as a function of centrality from peripheral to semi-central events and decreases in the most central collisions. 
The value of \acceff is 78\% in the 20-90\% centrality interval and 74\% in the 0-20\% interval, with centrality-independent systematic uncertainty of 5\%, as discussed below.

%
%
To evaluate the invariant yields (${\rm d} N / {\rm d} y$), the raw dimuon-triggered mass distribution must be normalised by the factor $\fnorm^i$, which is the inverse of the probability to observe a dimuon pair in a MB event for the centrality class $i$.
The value of $\fnorm^i$ is calculated in two different ways, by applying the dimuon selection criterion to MB events, and by the relative counting rate of the two triggers~\cite{Adam:2016rdg}.
The variation in  $\fnorm^i$ determined by these two methods is 0.5\% and  contributes to the systematic uncertainty.

%
%
The nuclear modification factor \Raa requires the determination of the collision centrality, which is typically quantified by the average number of nucleons participating in the interaction for a given centrality bin, \npart.
However, the rate of hard processes is known to scale with the average number of nucleon-nucleon collisions \ncoll.
The average centrality for hard processes is therefore presented as the average number of participant nucleons weighted by the number of collisions \npartNcoll.
\tab{tab:Centrality} shows the estimates of the average nuclear overlap function \taa, the number of participating nucleons $\langle \npart \rangle$ and the number of binary nucleon-nucleon collisions $\langle \ncoll \rangle$, which are obtained via a Glauber model fit of the signal amplitude in the two arrays of the V0 detector~\cite{Adam:2014qja,Adam:2015ptt}.
The resulting \npartNcoll is also shown.
The classification of the events in given centrality intervals has an associated uncertainty of 1.5 -- 2.3\% (centrality dependent), that was estimated by comparing the number of candidates selected by varying the centrality ranges by $\pm0.5\%$, to account for the centrality resolution~\cite{Adam:2014qja,Adam:2015ptt}.
\begin{table}[htb]
 \centering
 \begin{tabular}{c|cccc}
  \hline
  Centrality & \taa (\lumi[m]) & $\langle \npart \rangle$ & $\langle \ncoll \rangle$ & $\npartNcoll$ \\  \hline
  \centr{0}{90} &  $\hphantom{0}6.2\hphantom{0} \pm 0.2\hphantom{0}$ 		& $126 \pm 2$ 			&  $\hphantom{0}435 \pm \hphantom{0}41$ 		&  $ 263 \pm 3$ \\ 
  \centr{0}{20} &  $18.8\hphantom{0} \pm 0.6\hphantom{0} $ 							& $311 \pm 3$ 			&  $1318 \pm 130$				& $322 \pm 3$  \\
  \centr{20}{90} & $\hphantom{0}2.61 \pm 0.09$ 						& $\hphantom{0}73 \pm 1$ 		& $\hphantom{0}183 \pm \hphantom{0}15$ 		& $141 \pm 2$ \\ \hline
 \end{tabular}
 \caption{
 Values of the average nuclear overlap function, \taa, the number of participating nucleons, $\langle \npart \rangle$, and the number of binary nucleon-nucleon collisions, $\langle \ncoll \rangle$, for each centrality interval. The average number of participants as weighted by the average number of collisions, \npartNcoll, is also reported.}
 \label{tab:Centrality}
\end{table}

%
%
The sources of systematic uncertainties in the yields and \Raa are summarised in \tab{tab:syst}.
The systematic uncertainty in the tracking efficiency is 3\%, obtained from the comparison of the efficiency estimated in data and MC by exploiting the redundancy of the tracking chamber information~\cite{Abelev:2014ffa}.
The systematic uncertainty of the dimuon trigger efficiency is $1.5\%$, evaluated by propagating the uncertainty of the efficiency of the detection elements, which is estimated from data using the redundancy of the trigger chamber information. 
In addition, the choice of the $\chi^2$ cut used to match the tracker and trigger tracks introduces 1\% uncertainty, obtained from the difference between data and simulation when applying different $\chi^2$ cuts. The uncertainties in the track resolution and alignment are estimated by comparing the \acceff values obtained with two different simulations.
In the full simulation, the alignment is measured using the MILLIPEDE~\cite{Blobel:2002ax} package and the residual misalignment is taken into account.
In the fast simulation, the tracker response is based on a parameterisation of the measured resolution of the clusters associated with a track~\cite{Alice:2016wka}.
The resulting systematic uncertainty is 3.5\%. 

The total systematic uncertainty in the yield and \Raa are determined by summing in quadrature the  uncertainty from each source, listed in \tab{tab:syst}. All uncertainties except those due to \taa and the centrality bin boundaries are independent of collision centrality. 
Correlations in centrality or rapidity of the uncertainties of different sources are indicated in \tab{tab:syst}.
The relative systematic uncertainty in the proton-proton reference $\sigma_{\rm pp}$, which affects the \Raa, corresponds to 4.5\% and is estimated by varying the factorization and renormalisation scales and accounting for the uncertainties in the PDFs~\cite{Paukkunen:2010qg}.

\begin{table}[htb]
 \centering
 \begin{tabular}{l|c}
 \hline
 Source & Relative systematic uncertainty \\
 \hline \hline
 Background contamination & $<1.0\%$ \quad \phantom{0000} \\
 \hline
 Tracking efficiency & 		3.0\% \quad ($\star$) \\
 Trigger efficiency & 			1.5\% \quad ($\star$) \\
 Tracker/trigger matching & 	1.0\% \quad ($\star$) \\
 Alignment & 3.5\% \quad  ($\star$) \\
 \hline
 \fnorm & 0.5\% \;  ($\star \, \diamond$) \\
 $\sigma_{\rm pp}$ & 4.5\% \quad ($\star$) \\
 \hline
 \taa\ & 3.2 -- 3.5\% \quad  ($\diamond$) \\
 Centrality limits & 1.5 -- 2.3\% \quad ($\diamond$) \\
  \hline 
 \end{tabular}
 \caption{
Relative systematic uncertainties in the yields and \Raa. The ranges quoted for \taa and the centrality limits, represent the uncertainty variation with centrality. 
The centrality-dependent correlated uncertainties are marked by the symbol ($\star$), while the uncertainty sources that are correlated as a function of rapidity are indicated by ($\diamond$). 
}
 \label{tab:syst}
\end{table}


\section{Results}
\label{sec:results}

\begin{figure}[!htbp]
 \centering
 \includegraphics[width=0.45\textwidth]{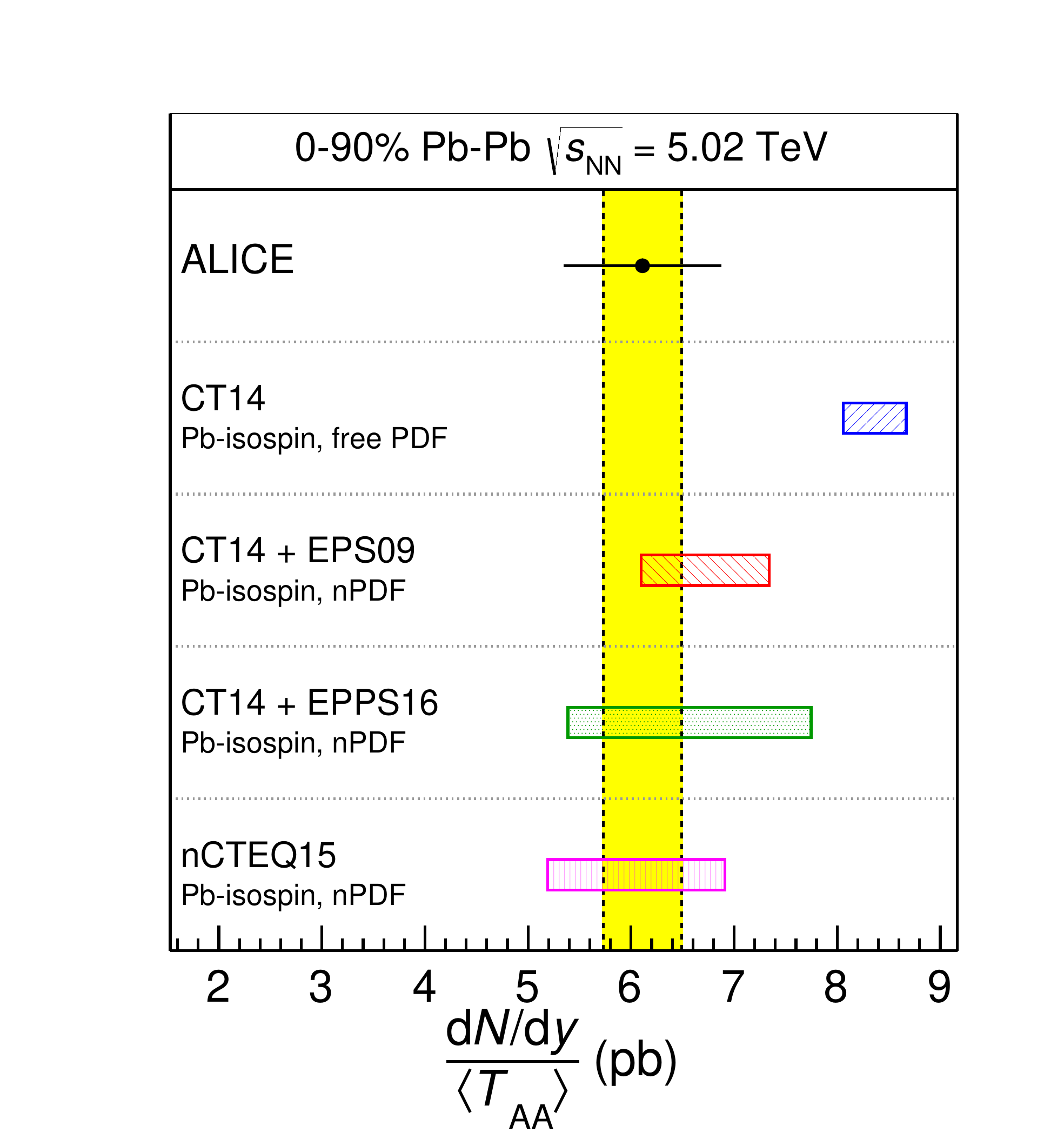}
 \caption{
Invariant yield of \dimuon from \zBoson production in $2.5<y<4.0$ divided by the average nuclear overlap function in the \centr{0}{90} centrality class, considering muons with $-4.0<\etamu<-2.5$ and $\ptmu>20$\GeVc. The horizontal solid line represents the statistical uncertainty of the measurement while the yellow filled band shows the systematic uncertainty. The result is compared to theoretical calculations with and without nuclear modification of the PDFs~\cite{Dulat:2015mca,Eskola:2009uj,Eskola:2016oht,Kovarik:2015cma,Kusina:2016fxy}. 
All model calculations incorporate PDFs or nPDFs determined by considering the isospin of the Pb-nucleus. 
 }
 \label{fig:Z_integrated}
\end{figure}

The invariant yield of \dimuon from \zBoson bosons in $2.5<y<4.0$, divided by \taa,
 is $6.11 \pm 0.76 \, {\rm (stat.)} \, \pm 0.38  \, {\rm (syst.)}$~pb for the \centr{0}{90} centrality interval. 
The comparison with theoretical calculations at NLO is shown in \fig{fig:Z_integrated}. 
The CT14~\cite{Dulat:2015mca} prediction utilises free proton and neutron PDFs, with relative weights to account for the isospin of the Pb nucleus.
The uncertainty on the model include the uncertainty on the NLO calculations and of the measurements considered in the PDF fit. 
The measured invariant yield deviates from the lower limit of this prediction by $2.3\sigma$. 
For the description of nuclear PDFs, two different approaches were considered. 
The standard approach evaluates the nPDF as the free PDF multiplied by a parameterisation of nuclear modifications.
The calculations obtained with the EPS09~\cite{Eskola:2009uj} and the more recent EPPS16~\cite{Eskola:2016oht} parameterisations are shown.
In the other approach, the nPDFs are obtained by fitting the nuclear data in a similar way as done for free proton data, but using a parameterisation that depends on the atomic mass of the nucleus.
The results obtained with the nCTEQ15 nuclear PDFs~\cite{Kovarik:2015cma,Kusina:2016fxy} are also presented.
The nPDF sets are characterised by their different approximations and by different input data included in the calculations (see~\ci{Eskola:2016oht} and references therein for details). 
Only the most recent EPPS16 parameterisation includes LHC jet, \wBoson and \zBoson data, although the \wBoson and \zBoson data provide only weak constraints on nPDFs at the current perturbative order of the calculation (NLO)~\cite{Armesto:2015lrg,Kusina:2016fxy}. 
In general, the nPDFs have larger uncertainties compared to the free proton PDFs, since they are less constrained from data.
CT14+EPS09 and CT14+EPPS16 estimates combine CT14 and EPS09 or EPPS16 uncertainties, whereas nCTEQ15 does a global study of the proton and nuclear measurement uncertainties included in the fit.
EPPS16 allows much more freedom for the flavour dependence of nPDFs than other current analyses, which results in larger uncertainties. 
All pQCD calculations shown in \fig{fig:Z_integrated} that use nPDFs describe the measurement well.

\begin{figure}[!htbp]
 \centering
\subfigure[]{
	\label{fig:ZYield_vs_y}
	\includegraphics[width=0.48\columnwidth]{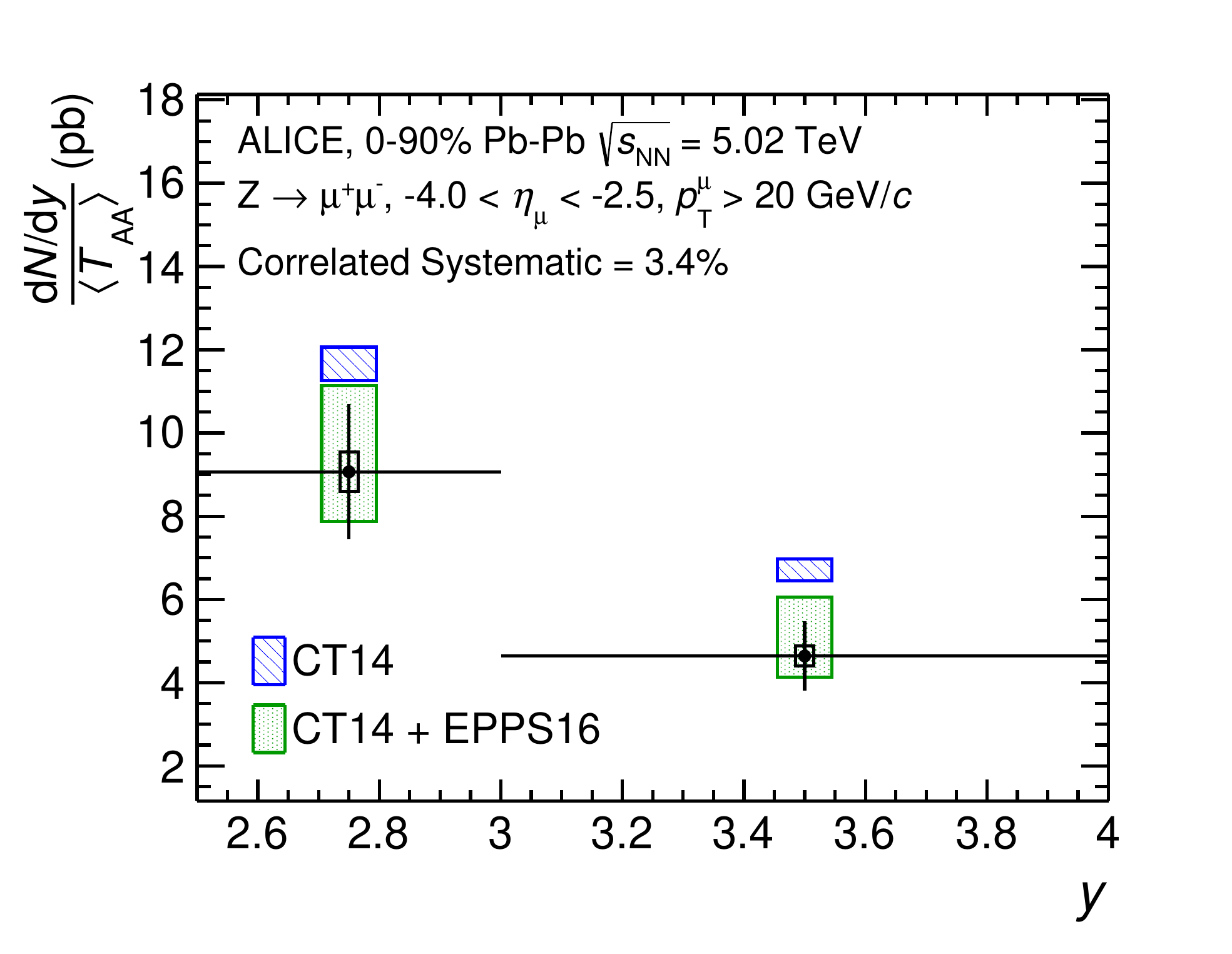}
	} 
\subfigure[]{
	\label{fig:ZRaa_vs_y}
	\includegraphics[width=0.48\columnwidth]{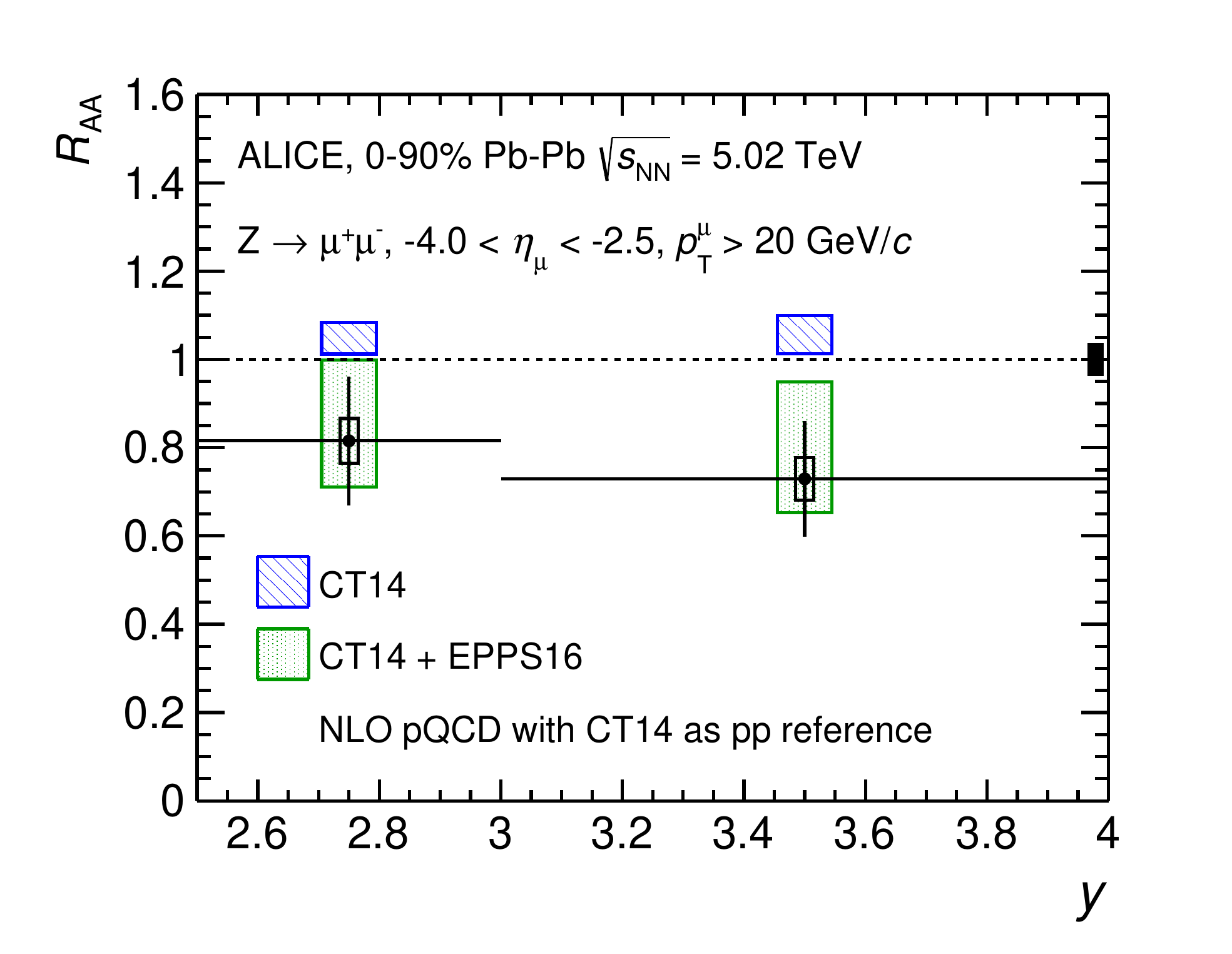}
	} 
\caption{
Invariant yield of \dimuon from \zBoson in $2.5<y<4.0$ divided by \taa (a) and nuclear modification factor (b) as a function of rapidity for \PbPb collisions at $\snn=5.02$~TeV, considering muons with $-4.0<\etamu<-2.5$ and $\ptmu>20$\GeVc. 
The vertical error bars are statistical only. The horizontal error bars display the measurement bin width, while the boxes represent the systematic uncertainties. 
The filled black box in panel (b), located at $\Raa=1$, shows the normalisation uncertainty. 
The results are compared to theoretical calculations with and without nuclear modification of the PDFs.
The filled blue boxes show the calculation using the CT14 PDF, while the green stippled boxes show the calculation using CT14 PDF with EPPS16 nPDF~\cite{Dulat:2015mca,Eskola:2016oht}.
All model calculations incorporate PDFs or nPDFs that account for the isospin of the Pb nucleus. 
}
\label{fig:Z_vs_y}
\end{figure}

%
%
The rapidity dependence of the \zBoson-boson invariant yields divided by \taa is shown in~\fig{fig:ZYield_vs_y}.
The results are compared to pQCD calculations using the CT14~\cite{Dulat:2015mca} PDF set both with (green filled box) and without (blue hatched box) the EPPS16~\cite{Eskola:2016oht} parameterisation of the nPDFs. 
In both cases, the Pb-isospin effect is modeled by combining the proton and neutron PDFs or nPDFs.
EPPS16 decreases the yields but does not have a strong influence on the rapidity dependence of the calculation.
The calculations that utilise vacuum PDFs overestimate data in the two rapidity intervals, whereas those that utilise nPDFs are in good agreement with data.

In this analysis, the ratio \Raa utilises a theoretically calculated reference cross section for pp  collisions~\cite{Paukkunen:2010qg}, which is $\sigma_{pp} = 11.92 \, \pm 0.43 $~pb.
The value of \Raa for the \centr{0}{90} centrality class is determined to be $0.77 \pm 0.10 \, {\rm (stat.)} \, \pm 0.06 \, {\rm (syst.)}$, deviating by $2.1 \sigma$ from unity.
The pQCD calculation using CT14~\cite{Dulat:2015mca} and considering only the isospin effects, finds $\Raa^{\rm CT14} = 1.052 \pm 0.038$.
The modification of the PDFs in nuclei results in a net reduction of the yields, and consequently in \Raa values lower than unity, with $\Raa^{\rm CT14+EPPS16} = 0.845 \pm 0.068$, in agreement with data.
The rapidity dependence of \Raa is presented in \fig{fig:ZRaa_vs_y}.
The values are smaller than unity, with a slight rapidity dependence.
The data are well-described by calculations including nPDFs (green filled boxes), while the calculations including only isospin effects (blue hatched boxes) tend to overestimate the measured values.

\begin{figure}[!!htbp]
\centering
\subfigure[]{
	\label{fig:ZYield_vs_Npart}
	\includegraphics[width=0.48\columnwidth]{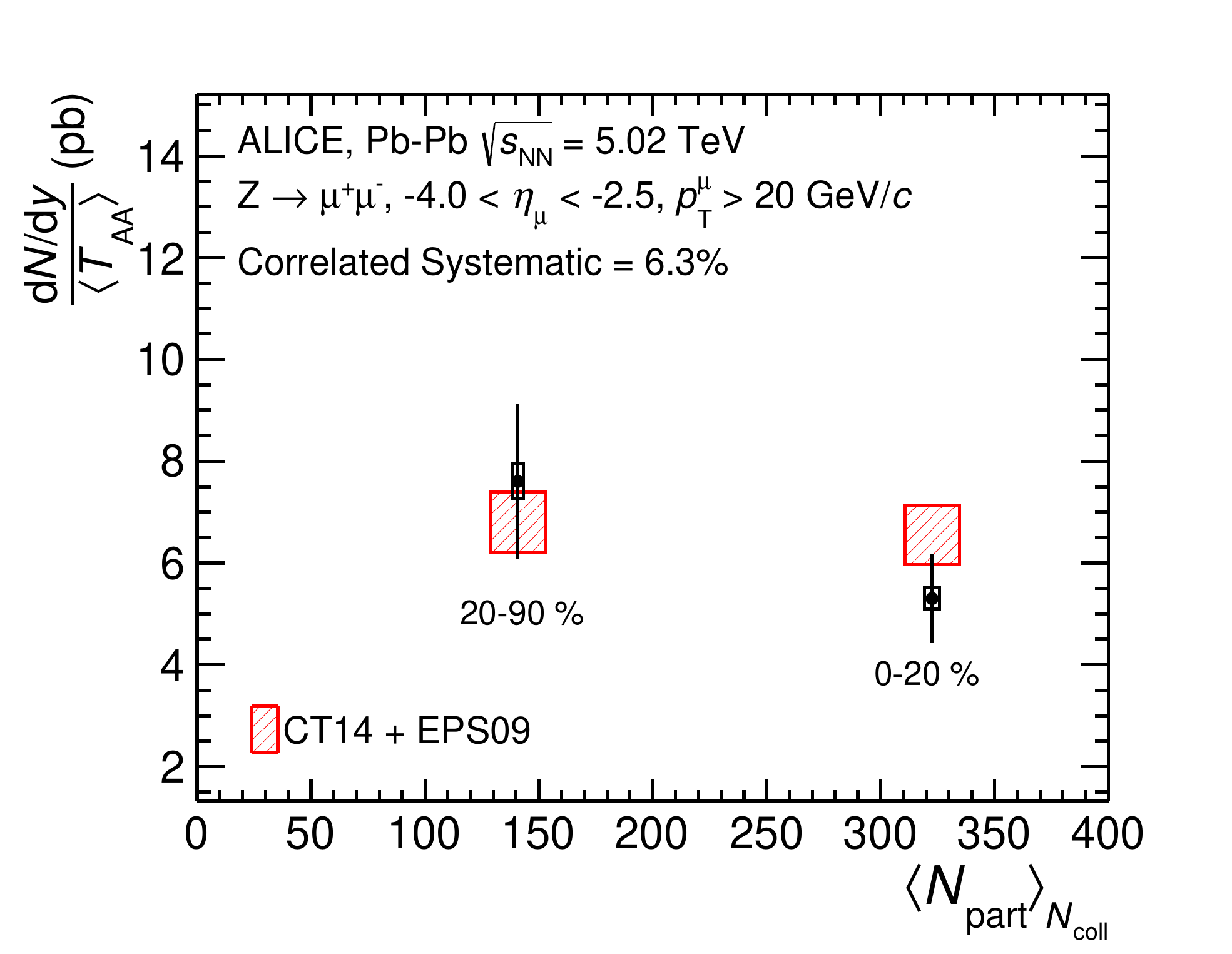}
	} 
\subfigure[ ]{
	\label{fig:ZRaa_vs_Npart}
	\includegraphics[width=0.48\columnwidth]{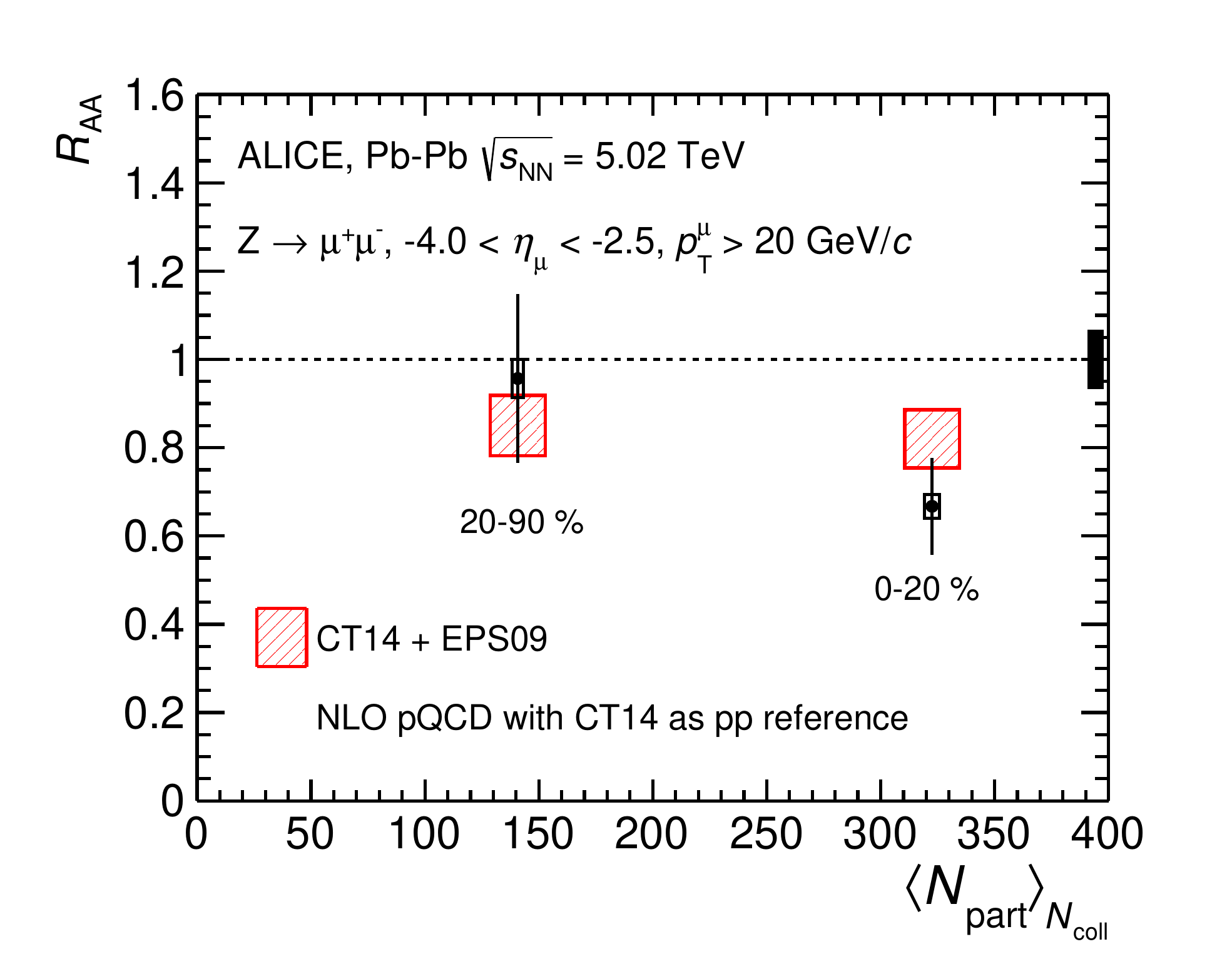}
	} 
\caption{
Invariant yield of \dimuon from \zBoson in $2.5<y<4.0$ divided by \taa (a) and nuclear modification factor (b) as a function of centrality (represented by \npartNcoll) for \PbPb collisions at $\snn=5.02$~TeV, considering muons with $-4.0<\etamu<-2.5$ and $\ptmu>20$\GeVc. 
The vertical error bars are statistical only, while the boxes represent the systematic uncertainties. 
The filled black box in panel (b), located at $\Raa=1$, shows the normalisation uncertainty. 
The results are compared to theoretical calculations with centrality-dependent nPDFs that account for the isospin of the Pb nucleus~\cite{Dulat:2015mca,Eskola:2009uj,Helenius:2012wd}. 
}
\label{fig:Z_vs_Npart}
\end{figure}

%
%
\zBoson-boson production is studied as a function of the collision centrality, expressed in terms of \npartNcoll\ as shown in \fig{fig:Z_vs_Npart}.
The value of \Raa is compatible with unity in peripheral collisions, with \Raa (\centr{20}{90}) = $0.96 \pm 0.19 \, \mbox{(stat.)} \, \pm 0.04 \, \mbox{(syst.)} \, \pm 0.06 \, \mbox{(corr. syst.)}$, while it is $2.6 \sigma$ smaller than unity in the central collisions, with \Raa (\centr{0}{20}) = $0.67 \pm 0.11 \, \mbox{(stat.)} \, \pm 0.03 \, \mbox{(syst.)} \, \pm 0.04 \, \mbox{(corr. syst.)}$.
The value for 0-20\% central collisions deviates from the predictions using vacuum PDFs ($\Raa^{\rm CT14}$) by $3\sigma$.
The data are compared to calculations including a centrality-dependent nuclear modification of the PDFs~\cite{Helenius:2012wd}, which describe the data within uncertainties.

\section{Conclusion}\label{sec:conclusion}
We have reported the first measurement of \zBoson-boson production at forward rapidities in \PbPb collisions at \snn = 5.02~TeV.
The invariant yields divided by the average nuclear overlap function are evaluated as a function of  rapidity and average number of participant nucleons weighted by the number of binary nucleon-nucleon collisions.
The corresponding values of the nuclear modification factor are estimated by dividing the measured yields in \PbPb collisions by the expected cross-section in pp collisions estimated with NLO pQCD calculations.
The value of \Raa is compatible with unity in the \centr{20}{90} centrality class (within large statistical uncertainty), whereas it is smaller than unity by 2.6 times the quadratic sum of the statistical and systematic uncertainties in the \centr{0}{20} most central collisions.
The results are well-described by the calculations that include modifications of the PDFs in nuclei.
In contrast, the calculations with vacuum PDFs overestimate the centrality-integrated \Raa by $2.3\sigma$ and \Raa in the \centr{0}{20} most central collisions by $3 \sigma$.

\newenvironment{acknowledgement}{\relax}{\relax}
\begin{acknowledgement}
\section*{Acknowledgements}
The ALICE Collaboration would like to thank H.~Paukkunen for providing the pQCD calculations with EPS09 and EPPS16, and F.~Lyonnet and A.~Kusina for providing the pQCD calculations with nCTEQ15. 

The ALICE Collaboration would like to thank all its engineers and technicians for their invaluable contributions to the construction of the experiment and the CERN accelerator teams for the outstanding performance of the LHC complex.
The ALICE Collaboration gratefully acknowledges the resources and support provided by all Grid centres and the Worldwide LHC Computing Grid (WLCG) collaboration.
The ALICE Collaboration acknowledges the following funding agencies for their support in building and running the ALICE detector:
A. I. Alikhanyan National Science Laboratory (Yerevan Physics Institute) Foundation (ANSL), State Committee of Science and World Federation of Scientists (WFS), Armenia;
Austrian Academy of Sciences and Nationalstiftung f\"{u}r Forschung, Technologie und Entwicklung, Austria;
Ministry of Communications and High Technologies, National Nuclear Research Center, Azerbaijan;
Conselho Nacional de Desenvolvimento Cient\'{\i}fico e Tecnol\'{o}gico (CNPq), Universidade Federal do Rio Grande do Sul (UFRGS), Financiadora de Estudos e Projetos (Finep) and Funda\c{c}\~{a}o de Amparo \`{a} Pesquisa do Estado de S\~{a}o Paulo (FAPESP), Brazil;
Ministry of Science \& Technology of China (MSTC), National Natural Science Foundation of China (NSFC) and Ministry of Education of China (MOEC) , China;
Ministry of Science, Education and Sport and Croatian Science Foundation, Croatia;
Ministry of Education, Youth and Sports of the Czech Republic, Czech Republic;
The Danish Council for Independent Research | Natural Sciences, the Carlsberg Foundation and Danish National Research Foundation (DNRF), Denmark;
Helsinki Institute of Physics (HIP), Finland;
Commissariat \`{a} l'Energie Atomique (CEA) and Institut National de Physique Nucl\'{e}aire et de Physique des Particules (IN2P3) and Centre National de la Recherche Scientifique (CNRS), France;
Bundesministerium f\"{u}r Bildung, Wissenschaft, Forschung und Technologie (BMBF) and GSI Helmholtzzentrum f\"{u}r Schwerionenforschung GmbH, Germany;
General Secretariat for Research and Technology, Ministry of Education, Research and Religions, Greece;
National Research, Development and Innovation Office, Hungary;
Department of Atomic Energy Government of India (DAE), Department of Science and Technology, Government of India (DST), University Grants Commission, Government of India (UGC) and Council of Scientific and Industrial Research (CSIR), India;
Indonesian Institute of Science, Indonesia;
Centro Fermi - Museo Storico della Fisica e Centro Studi e Ricerche Enrico Fermi and Istituto Nazionale di Fisica Nucleare (INFN), Italy;
Institute for Innovative Science and Technology , Nagasaki Institute of Applied Science (IIST), Japan Society for the Promotion of Science (JSPS) KAKENHI and Japanese Ministry of Education, Culture, Sports, Science and Technology (MEXT), Japan;
Consejo Nacional de Ciencia (CONACYT) y Tecnolog\'{i}a, through Fondo de Cooperaci\'{o}n Internacional en Ciencia y Tecnolog\'{i}a (FONCICYT) and Direcci\'{o}n General de Asuntos del Personal Academico (DGAPA), Mexico;
Nederlandse Organisatie voor Wetenschappelijk Onderzoek (NWO), Netherlands;
The Research Council of Norway, Norway;
Commission on Science and Technology for Sustainable Development in the South (COMSATS), Pakistan;
Pontificia Universidad Cat\'{o}lica del Per\'{u}, Peru;
Ministry of Science and Higher Education and National Science Centre, Poland;
Korea Institute of Science and Technology Information and National Research Foundation of Korea (NRF), Republic of Korea;
Ministry of Education and Scientific Research, Institute of Atomic Physics and Romanian National Agency for Science, Technology and Innovation, Romania;
Joint Institute for Nuclear Research (JINR), Ministry of Education and Science of the Russian Federation and National Research Centre Kurchatov Institute, Russia;
Ministry of Education, Science, Research and Sport of the Slovak Republic, Slovakia;
National Research Foundation of South Africa, South Africa;
Centro de Aplicaciones Tecnol\'{o}gicas y Desarrollo Nuclear (CEADEN), Cubaenerg\'{\i}a, Cuba and Centro de Investigaciones Energ\'{e}ticas, Medioambientales y Tecnol\'{o}gicas (CIEMAT), Spain;
Swedish Research Council (VR) and Knut \& Alice Wallenberg Foundation (KAW), Sweden;
European Organization for Nuclear Research, Switzerland;
National Science and Technology Development Agency (NSDTA), Suranaree University of Technology (SUT) and Office of the Higher Education Commission under NRU project of Thailand, Thailand;
Turkish Atomic Energy Agency (TAEK), Turkey;
National Academy of  Sciences of Ukraine, Ukraine;
Science and Technology Facilities Council (STFC), United Kingdom;
National Science Foundation of the United States of America (NSF) and United States Department of Energy, Office of Nuclear Physics (DOE NP), United States of America.
\end{acknowledgement}


\bibliographystyle{utphys}   
\bibliography{biblio}

\providecommand{\href}[2]{#2}\begingroup\raggedright\begin{thebibliography}{10}

\bibitem{Peigne:2007sd}
S.~Peigne and A.~Peshier, ``{Collisional Energy Loss of a Fast Muon in a Hot
  QED Plasma},'' \href{http://dx.doi.org/10.1103/PhysRevD.77.014015}{{\em Phys.
  Rev.} {\bfseries D77} (2008) 014015},
\href{http://arxiv.org/abs/0710.1266}{{\ttfamily arXiv:0710.1266 [hep-ph]}}.

\bibitem{Kartvelishvili:1995fr}
V.~Kartvelishvili, R.~Kvatadze, and R.~Shanidze, ``{On Z and Z + jet production
  in heavy ion collisions},''
  \href{http://dx.doi.org/10.1016/0370-2693(95)00865-I}{{\em Phys. Lett.}
  {\bfseries B356} (1995) 589--594},
\href{http://arxiv.org/abs/hep-ph/9505418}{{\ttfamily arXiv:hep-ph/9505418
  [hep-ph]}}.

\bibitem{ConesadelValle:2009vp}
Z.~Conesa~del Valle, ``{Vector bosons in heavy-ion collisions at the LHC},''
  \href{http://dx.doi.org/10.1140/epjc/s10052-009-0980-8}{{\em Eur. Phys. J.}
  {\bfseries C61} (2009) 729--733},
\href{http://arxiv.org/abs/0903.1432}{{\ttfamily arXiv:0903.1432 [hep-ex]}}.

\bibitem{Albajar:1987yz}
{\bfseries UA1} Collaboration, C.~Albajar {\em et~al.}, ``{Intermediate Vector
  Boson Cross-Sections at the CERN Super Proton Synchrotron Collider and the
  Number of Neutrino Types},''
\href{http://dx.doi.org/10.1016/0370-2693(87)91510-3}{{\em Phys. Lett.}
  {\bfseries B198} (1987) 271}.

\bibitem{Alitti:1991dm}
{\bfseries UA2} Collaboration, J.~Alitti {\em et~al.}, ``{A Measurement of the
  $W$ and $Z$ production cross-sections and a determination of $\Gamma_{W}$ at
  the CERN $\bar{p} p$ collider},''
\href{http://dx.doi.org/10.1016/0370-2693(92)90333-Y}{{\em Phys. Lett.}
  {\bfseries B276} (1992) 365--374}.

\bibitem{Abe:1995bm}
{\bfseries CDF} Collaboration, F.~Abe {\em et~al.}, ``{Measurement of $\sigma
  \cdot B (W \to e \nu)$ and $\sigma \cdot B(Z^0 \to e^+e^-)$ in $p\bar{p}$
  collisions at $\sqrt{s}=1.8$ TeV},''
  \href{http://dx.doi.org/10.1103/PhysRevLett.76.3070}{{\em Phys. Rev. Lett.}
  {\bfseries 76} (1996) 3070--3075},
\href{http://arxiv.org/abs/hep-ex/9509010}{{\ttfamily arXiv:hep-ex/9509010
  [hep-ex]}}.

\bibitem{Abbott:1999tt}
{\bfseries D0} Collaboration, B.~Abbott {\em et~al.}, ``{Extraction of the
  width of the $W$ boson from measurements of $\sigma(p\bar{p} \to W + X)
  \times B(W \to e \nu)$ and $\sigma(p\bar{p} \to Z + X) \times B(Z \to e e)$
  and their ratio},'' \href{http://dx.doi.org/10.1103/PhysRevD.61.072001}{{\em
  Phys. Rev.} {\bfseries D61} (2000) 072001},
\href{http://arxiv.org/abs/hep-ex/9906025}{{\ttfamily arXiv:hep-ex/9906025
  [hep-ex]}}.

\bibitem{Abulencia:2005ix}
{\bfseries CDF} Collaboration, A.~Abulencia {\em et~al.}, ``{Measurements of
  inclusive W and Z cross sections in $p\overline{p}$ collisions at $\sqrt{s}$
  = 1.96 TeV},'' \href{http://dx.doi.org/10.1088/0954-3899/34/12/001}{{\em J.
  Phys.} {\bfseries G34} (2007) 2457--2544},
\href{http://arxiv.org/abs/hep-ex/0508029}{{\ttfamily arXiv:hep-ex/0508029
  [hep-ex]}}.

\bibitem{Aad:2010yt}
{\bfseries ATLAS} Collaboration, G.~Aad {\em et~al.}, ``{Measurement of the $W
  \to \ell\nu$ and $Z/\gamma^* \to \ell\ell$ production cross sections in
  proton-proton collisions at $\sqrt{s} = 7$ TeV with the ATLAS detector},''
  \href{http://dx.doi.org/10.1007/JHEP12(2010)060}{{\em JHEP} {\bfseries 12}
  (2010) 060},
\href{http://arxiv.org/abs/1010.2130}{{\ttfamily arXiv:1010.2130 [hep-ex]}}.

\bibitem{Aad:2011dm}
{\bfseries ATLAS} Collaboration, G.~Aad {\em et~al.}, ``{Measurement of the
  inclusive $W^\pm$ and $Z/\gamma^*$ cross sections in the electron and muon
  decay channels in $pp$ collisions at $\sqrt{s}=7$ TeV with the ATLAS
  detector},'' \href{http://dx.doi.org/10.1103/PhysRevD.85.072004}{{\em Phys.
  Rev.} {\bfseries D85} (2012) 072004},
\href{http://arxiv.org/abs/1109.5141}{{\ttfamily arXiv:1109.5141 [hep-ex]}}.

\bibitem{Aad:2016naf}
{\bfseries ATLAS} Collaboration, G.~Aad {\em et~al.}, ``{Measurement of
  $W^{\pm}$ and $Z$-boson production cross sections in $pp$ collisions at
  $\sqrt{s}=13$ TeV with the ATLAS detector},''
  \href{http://dx.doi.org/10.1016/j.physletb.2016.06.023}{{\em Phys. Lett.}
  {\bfseries B759} (2016) 601--621},
\href{http://arxiv.org/abs/1603.09222}{{\ttfamily arXiv:1603.09222 [hep-ex]}}.

\bibitem{CMS:2011aa}
{\bfseries CMS} Collaboration, S.~Chatrchyan {\em et~al.}, ``{Measurement of
  the Inclusive $W$ and $Z$ Production Cross Sections in $pp$ Collisions at
  $\sqrt{s}=7$ TeV},'' \href{http://dx.doi.org/10.1007/JHEP10(2011)132}{{\em
  JHEP} {\bfseries 10} (2011) 132},
\href{http://arxiv.org/abs/1107.4789}{{\ttfamily arXiv:1107.4789 [hep-ex]}}.

\bibitem{Chatrchyan:2014mua}
{\bfseries CMS} Collaboration, S.~Chatrchyan {\em et~al.}, ``{Measurement of
  inclusive W and Z boson production cross sections in pp collisions at
  $\sqrt{s}$ = 8 TeV},''
  \href{http://dx.doi.org/10.1103/PhysRevLett.112.191802}{{\em Phys. Rev.
  Lett.} {\bfseries 112} (2014) 191802},
\href{http://arxiv.org/abs/1402.0923}{{\ttfamily arXiv:1402.0923 [hep-ex]}}.

\bibitem{Aaij:2015gna}
{\bfseries LHCb} Collaboration, R.~Aaij {\em et~al.}, ``{Measurement of the
  forward $Z$ boson production cross-section in $pp$ collisions at $\sqrt{s}=7$
  TeV},'' \href{http://dx.doi.org/10.1007/JHEP08(2015)039}{{\em JHEP}
  {\bfseries 08} (2015) 039},
\href{http://arxiv.org/abs/1505.07024}{{\ttfamily arXiv:1505.07024 [hep-ex]}}.

\bibitem{Aaij:2015zlq}
{\bfseries LHCb} Collaboration, R.~Aaij {\em et~al.}, ``{Measurement of forward
  W and Z boson production in $pp$ collisions at $ \sqrt{s}=8 $ TeV},''
  \href{http://dx.doi.org/10.1007/JHEP01(2016)155}{{\em JHEP} {\bfseries 01}
  (2016) 155},
\href{http://arxiv.org/abs/1511.08039}{{\ttfamily arXiv:1511.08039 [hep-ex]}}.

\bibitem{Martin:1999ww}
A.~D. Martin, R.~G. Roberts, W.~J. Stirling, and R.~S. Thorne, ``{Parton
  distributions and the LHC: $W$ and $Z$ production},''
  \href{http://dx.doi.org/10.1007/s100520050740}{{\em Eur. Phys. J.} {\bfseries
  C14} (2000) 133--145},
\href{http://arxiv.org/abs/hep-ph/9907231}{{\ttfamily arXiv:hep-ph/9907231
  [hep-ph]}}.

\bibitem{Baur:1998kt}
U.~Baur, S.~Keller, and D.~Wackeroth, ``{Electroweak radiative corrections to
  $W$ boson production in hadronic collisions},''
  \href{http://dx.doi.org/10.1103/PhysRevD.59.013002}{{\em Phys. Rev.}
  {\bfseries D59} (1999) 013002},
\href{http://arxiv.org/abs/hep-ph/9807417}{{\ttfamily arXiv:hep-ph/9807417
  [hep-ph]}}.

\bibitem{Rojo:2015acz}
J.~Rojo {\em et~al.}, ``{The PDF4LHC report on PDFs and LHC data: Results from
  Run I and preparation for Run II},''
  \href{http://dx.doi.org/10.1088/0954-3899/42/10/103103}{{\em J. Phys.}
  {\bfseries G42} (2015) 103103},
\href{http://arxiv.org/abs/1507.00556}{{\ttfamily arXiv:1507.00556 [hep-ph]}}.

\bibitem{Butterworth:2015oua}
J.~Butterworth {\em et~al.}, ``{PDF4LHC recommendations for LHC Run II},''
  \href{http://dx.doi.org/10.1088/0954-3899/43/2/023001}{{\em J. Phys.}
  {\bfseries G43} (2016) 023001},
\href{http://arxiv.org/abs/1510.03865}{{\ttfamily arXiv:1510.03865 [hep-ph]}}.

\bibitem{Paukkunen:2010qg}
H.~Paukkunen and C.~A. Salgado, ``{Constraints for the nuclear parton
  distributions from Z and W production at the LHC},''
  \href{http://dx.doi.org/10.1007/JHEP03(2011)071}{{\em JHEP} {\bfseries 1103}
  (2011) 071},
\href{http://arxiv.org/abs/1010.5392}{{\ttfamily arXiv:1010.5392 [hep-ph]}}.

\bibitem{Kusina:2016fxy}
A.~Kusina, F.~Lyonnet, D.~B. Clark, E.~Godat, T.~Jezo, K.~Kovarik, F.~I.
  Olness, I.~Schienbein, and J.~Y. Yu, ``{Vector boson production in
  proton-lead and lead-lead collisions at the LHC and its impact on nCTEQ15
  PDFs},''
\href{http://arxiv.org/abs/1610.02925}{{\ttfamily arXiv:1610.02925 [nucl-th]}}.

\bibitem{Eskola:2016oht}
K.~J. Eskola, P.~Paakkinen, H.~Paukkunen, and C.~A. Salgado, ``{EPPS16: Nuclear
  parton distributions with LHC data},''
  \href{http://dx.doi.org/10.1140/epjc/s10052-017-4725-9}{{\em Eur. Phys. J.}
  {\bfseries C77} (2017) 163},
\href{http://arxiv.org/abs/1612.05741}{{\ttfamily arXiv:1612.05741 [hep-ph]}}.

\bibitem{deFlorian:2003qf}
D.~de~Florian and R.~Sassot, ``{Nuclear parton distributions at next-to-leading
  order},'' \href{http://dx.doi.org/10.1103/PhysRevD.69.074028}{{\em Phys.
  Rev.} {\bfseries D69} (2004) 074028},
\href{http://arxiv.org/abs/hep-ph/0311227}{{\ttfamily arXiv:hep-ph/0311227
  [hep-ph]}}.

\bibitem{Hirai:2007sx}
M.~Hirai, S.~Kumano, and T.~H. Nagai, ``{Determination of nuclear parton
  distribution functions and their uncertainties in next-to-leading order},''
  \href{http://dx.doi.org/10.1103/PhysRevC.76.065207}{{\em Phys. Rev.}
  {\bfseries C76} (2007) 065207},
\href{http://arxiv.org/abs/0709.3038}{{\ttfamily arXiv:0709.3038 [hep-ph]}}.

\bibitem{AtashbarTehrani:2012xh}
S.~Atashbar~Tehrani, ``{Nuclear parton densities and their uncertainties at the
  next-to-leading order},''
\href{http://dx.doi.org/10.1103/PhysRevC.86.064301}{{\em Phys. Rev.} {\bfseries
  C86} (2012) 064301}.

\bibitem{Khanpour:2016pph}
H.~Khanpour and S.~Atashbar~Tehrani, ``{Global Analysis of Nuclear Parton
  Distribution Functions and Their Uncertainties at Next-to-Next-to-Leading
  Order},'' \href{http://dx.doi.org/10.1103/PhysRevD.93.014026}{{\em Phys.
  Rev.} {\bfseries D93} (2016) 014026},
\href{http://arxiv.org/abs/1601.00939}{{\ttfamily arXiv:1601.00939 [hep-ph]}}.

\bibitem{Vogt:2000hp}
R.~Vogt, ``{Shadowing effects on vector boson production},''
  \href{http://dx.doi.org/10.1103/PhysRevC.64.044901}{{\em Phys. Rev.}
  {\bfseries C64} (2001) 044901},
\href{http://arxiv.org/abs/hep-ph/0011242}{{\ttfamily arXiv:hep-ph/0011242
  [hep-ph]}}.

\bibitem{Neufeld:2010dz}
R.~B. Neufeld, I.~Vitev, and B.-W. Zhang, ``{A possible determination of the
  quark radiation length in cold nuclear matter},''
  \href{http://dx.doi.org/10.1016/j.physletb.2011.09.045}{{\em Phys. Lett.}
  {\bfseries B704} (2011) 590--595},
\href{http://arxiv.org/abs/1010.3708}{{\ttfamily arXiv:1010.3708 [hep-ph]}}.

\bibitem{Aad:2015gta}
{\bfseries ATLAS} Collaboration, G.~Aad {\em et~al.}, ``{$Z$ boson production
  in $p+$Pb collisions at $\sqrt{s_{\mathrm{NN}}}=5.02$ TeV measured with the
  ATLAS detector},'' \href{http://dx.doi.org/10.1103/PhysRevC.92.044915}{{\em
  Phys. Rev.} {\bfseries C92} (2015) 044915},
\href{http://arxiv.org/abs/1507.06232}{{\ttfamily arXiv:1507.06232 [hep-ex]}}.

\bibitem{Khachatryan:2015pzs}
{\bfseries CMS} Collaboration, V.~Khachatryan {\em et~al.}, ``{Study of Z boson
  production in pPb collisions at $\sqrt{s_{\mathrm{NN}}}=5.02$ TeV},''
  \href{http://dx.doi.org/10.1016/j.physletb.2016.05.044}{{\em Phys. Lett.}
  {\bfseries B759} (2016) 36--57},
\href{http://arxiv.org/abs/1512.06461}{{\ttfamily arXiv:1512.06461 [hep-ex]}}.

\bibitem{Khachatryan:2015hha}
{\bfseries CMS} Collaboration, V.~Khachatryan {\em et~al.}, ``{Study of W boson
  production in pPb collisions at $\sqrt{s_{\mathrm{NN}}} =$ 5.02 TeV},''
  \href{http://dx.doi.org/10.1016/j.physletb.2015.09.057}{{\em Phys. Lett.}
  {\bfseries B750} (2015) 565--586},
\href{http://arxiv.org/abs/1503.05825}{{\ttfamily arXiv:1503.05825 [nucl-ex]}}.

\bibitem{Alice:2016wka}
{\bfseries ALICE} Collaboration, J.~Adam {\em et~al.}, ``{W and Z boson
  production in p-Pb collisions at $\sqrt{s_{\rm NN}}$ = 5.02 TeV},''
  \href{http://dx.doi.org/10.1007/JHEP02(2017)077}{{\em JHEP} {\bfseries 02}
  (2017) 077},
\href{http://arxiv.org/abs/1611.03002}{{\ttfamily arXiv:1611.03002 [nucl-ex]}}.

\bibitem{Aaij:2014pvu}
{\bfseries LHCb} Collaboration, R.~Aaij {\em et~al.}, ``{Observation of $Z$
  production in proton-lead collisions at LHCb},''
  \href{http://dx.doi.org/10.1007/JHEP09(2014)030}{{\em JHEP} {\bfseries 1409}
  (2014) 030},
\href{http://arxiv.org/abs/1406.2885}{{\ttfamily arXiv:1406.2885 [hep-ex]}}.

\bibitem{Gavin:2010az}
R.~Gavin, Y.~Li, F.~Petriello, and S.~Quackenbush, ``{FEWZ 2.0: A code for
  hadronic Z production at Next-to-Next-to-Leading order},''
  \href{http://dx.doi.org/10.1016/j.cpc.2011.06.008}{{\em Comput. Phys.
  Commun.} {\bfseries 182} (2011) 2388--2403},
\href{http://arxiv.org/abs/1011.3540}{{\ttfamily arXiv:1011.3540 [hep-ph]}}.

\bibitem{Aad:2014bha}
{\bfseries ATLAS} Collaboration, G.~Aad {\em et~al.}, ``{Measurement of the
  production and lepton charge asymmetry of $W$ bosons in Pb+Pb collisions at
  $\sqrt{s_{\mathrm {NN}}}=2.76$~TeV with the ATLAS detector},''
  \href{http://dx.doi.org/10.1140/epjc/s10052-014-3231-6}{{\em Eur.Phys.J.}
  {\bfseries C75} (2015) 23},
\href{http://arxiv.org/abs/1408.4674}{{\ttfamily arXiv:1408.4674 [hep-ex]}}.

\bibitem{Aad:2012ew}
{\bfseries ATLAS} Collaboration, G.~Aad {\em et~al.}, ``{Measurement of $Z$
  boson Production in Pb+Pb Collisions at $\sqrt{s_\mathrm{NN}}=2.76$ TeV with
  the ATLAS Detector},''
  \href{http://dx.doi.org/10.1103/PhysRevLett.110.022301}{{\em Phys.Rev.Lett.}
  {\bfseries 110} (2013) 022301},
\href{http://arxiv.org/abs/1210.6486}{{\ttfamily arXiv:1210.6486 [hep-ex]}}.

\bibitem{Chatrchyan:2014csa}
{\bfseries CMS} Collaboration, S.~Chatrchyan {\em et~al.}, ``{Study of Z
  production in PbPb and pp collisions at $ \sqrt{s_{\mathrm{NN}}}=2.76 $ TeV
  in the dimuon and dielectron decay channels},''
  \href{http://dx.doi.org/10.1007/JHEP03(2015)022}{{\em JHEP} {\bfseries 1503}
  (2015) 022},
\href{http://arxiv.org/abs/1410.4825}{{\ttfamily arXiv:1410.4825 [nucl-ex]}}.

\bibitem{Chatrchyan:2012nt}
{\bfseries CMS} Collaboration, S.~Chatrchyan {\em et~al.}, ``{Study of $W$
  boson production in PbPb and $pp$ collisions at $\sqrt{s_\mathrm{NN}}=2.76$
  TeV},'' \href{http://dx.doi.org/10.1016/j.physletb.2012.07.025}{{\em
  Phys.Lett.} {\bfseries B715} (2012) 66--87},
\href{http://arxiv.org/abs/1205.6334}{{\ttfamily arXiv:1205.6334 [nucl-ex]}}.

\bibitem{ATLAS:2017zkv}
{\bfseries ATLAS} Collaboration, ``{$Z$ boson production in Pb+Pb collisions
  at$\sqrt{s_{\mathrm{NN}}}=5.02$ TeV with the ATLAS detector at the LHC},''
{\em ATLAS-CONF-2017-010} (2017) .

\bibitem{Miller:2007ri}
M.~L. Miller, K.~Reygers, S.~J. Sanders, and P.~Steinberg, ``{Glauber modeling
  in high energy nuclear collisions},''
  \href{http://dx.doi.org/10.1146/annurev.nucl.57.090506.123020}{{\em Ann. Rev.
  Nucl. Part. Sci.} {\bfseries 57} (2007) 205--243},
\href{http://arxiv.org/abs/nucl-ex/0701025}{{\ttfamily arXiv:nucl-ex/0701025
  [nucl-ex]}}.

\bibitem{Aamodt:2008zz}
{\bfseries ALICE} Collaboration, K.~Aamodt {\em et~al.}, ``{The ALICE
  experiment at the CERN LHC},''
\href{http://dx.doi.org/10.1088/1748-0221/3/08/S08002}{{\em JINST} {\bfseries
  3} (2008) S08002}.

\bibitem{ALICE:2012aa}
{\bfseries ALICE} Collaboration, B.~Abelev {\em et~al.}, ``{Measurement of the
  Cross Section for Electromagnetic Dissociation with Neutron Emission in Pb-Pb
  Collisions at $\sqrt{s_{NN}}$ = 2.76 TeV},''
  \href{http://dx.doi.org/10.1103/PhysRevLett.109.252302}{{\em Phys. Rev.
  Lett.} {\bfseries 109} (2012) 252302},
\href{http://arxiv.org/abs/1203.2436}{{\ttfamily arXiv:1203.2436 [nucl-ex]}}.

\bibitem{Aamodt:2011gj}
{\bfseries ALICE} Collaboration, K.~Aamodt {\em et~al.}, ``{Rapidity and
  transverse momentum dependence of inclusive J$/\psi$ production in $pp$
  collisions at $\sqrt{s} = 7$ TeV},''
  \href{http://dx.doi.org/10.1016/j.physletb.2011.09.054}{{\em Phys. Lett.}
  {\bfseries B704} (2011) 442--455},
  \href{http://arxiv.org/abs/1105.0380}{{\ttfamily arXiv:1105.0380 [hep-ex]}}.
[Erratum: \href{http://dx.doi.org/10.1016/j.physletb.2012.10.060}{Phys.
  Lett.B718,692(2012)}].

\bibitem{Alioli:2008gx}
S.~Alioli, P.~Nason, C.~Oleari, and E.~Re, ``{NLO vector-boson production
  matched with shower in POWHEG},''
  \href{http://dx.doi.org/10.1088/1126-6708/2008/07/060}{{\em JHEP} {\bfseries
  07} (2008) 060},
\href{http://arxiv.org/abs/0805.4802}{{\ttfamily arXiv:0805.4802 [hep-ph]}}.

\bibitem{Sjostrand:2006za}
T.~Sjostrand, S.~Mrenna, and P.~Z. Skands, ``{PYTHIA 6.4 Physics and Manual},''
  \href{http://dx.doi.org/10.1088/1126-6708/2006/05/026}{{\em JHEP} {\bfseries
  05} (2006) 026},
\href{http://arxiv.org/abs/hep-ph/0603175}{{\ttfamily arXiv:hep-ph/0603175
  [hep-ph]}}.

\bibitem{Brun:1994aa}
R.~Brun, F.~Carminati, and S.~Giani, ``{GEANT Detector Description and
  Simulation Tool},''
{\em CERN-W-5013} (1994) .

\bibitem{Chatrchyan:2011wt}
{\bfseries CMS} Collaboration, S.~Chatrchyan {\em et~al.}, ``{Measurement of
  the Rapidity and Transverse Momentum Distributions of $Z$ Bosons in $pp$
  Collisions at $\sqrt{s}=7$ TeV},''
  \href{http://dx.doi.org/10.1103/PhysRevD.85.032002}{{\em Phys. Rev.}
  {\bfseries D85} (2012) 032002},
\href{http://arxiv.org/abs/1110.4973}{{\ttfamily arXiv:1110.4973 [hep-ex]}}.

\bibitem{Adam:2016rdg}
{\bfseries ALICE} Collaboration, J.~Adam {\em et~al.}, ``{J/$\psi$ suppression
  at forward rapidity in Pb-Pb collisions at $\sqrt{s_{{\rm NN}}} = 5.02$
  TeV},'' \href{http://dx.doi.org/10.1016/j.physletb.2016.12.064}{{\em Phys.
  Lett.} {\bfseries B766} (2017) 212--224},
\href{http://arxiv.org/abs/1606.08197}{{\ttfamily arXiv:1606.08197 [nucl-ex]}}.

\bibitem{Adam:2014qja}
{\bfseries ALICE} Collaboration, J.~Adam {\em et~al.}, ``{Centrality dependence
  of particle production in p-Pb collisions at $\sqrt{s_{\rm NN} }= 5.02$
  TeV},'' \href{http://dx.doi.org/10.1103/PhysRevC.91.064905}{{\em Phys.Rev.}
  {\bfseries C91} (2015) 064905},
\href{http://arxiv.org/abs/1412.6828}{{\ttfamily arXiv:1412.6828 [nucl-ex]}}.

\bibitem{Adam:2015ptt}
{\bfseries ALICE} Collaboration, J.~Adam {\em et~al.}, ``{Centrality dependence
  of the charged-particle multiplicity density at midrapidity in Pb-Pb
  collisions at $\sqrt{s_{\rm NN}}$ = 5.02 TeV},''
  \href{http://dx.doi.org/10.1103/PhysRevLett.116.222302}{{\em Phys. Rev.
  Lett.} {\bfseries 116} (2016) 222302},
\href{http://arxiv.org/abs/1512.06104}{{\ttfamily arXiv:1512.06104 [nucl-ex]}}.

\bibitem{Abelev:2014ffa}
{\bfseries ALICE} Collaboration, B.~B. Abelev {\em et~al.}, ``{Performance of
  the ALICE Experiment at the CERN LHC},''
  \href{http://dx.doi.org/10.1142/S0217751X14300440}{{\em Int. J. Mod. Phys.}
  {\bfseries A29} (2014) 1430044},
\href{http://arxiv.org/abs/1402.4476}{{\ttfamily arXiv:1402.4476 [nucl-ex]}}.

\bibitem{Blobel:2002ax}
V.~Blobel and C.~Kleinwort, ``{A New method for the high precision alignment of
  track detectors},'' in {\em {Advanced statistical techniques in particle
  physics. Proceedings, Conference, Durham, UK, March 18-22, 2002}},
  pp.~URL--STR(9).
\newblock 2002.
\newblock \href{http://arxiv.org/abs/hep-ex/0208021}{{\ttfamily
  arXiv:hep-ex/0208021 [hep-ex]}}.
\newblock
\url{http://www.ippp.dur.ac.uk/Workshops/02/statistics/proceedings//blobel1.pdf}.
\newblock

\bibitem{Dulat:2015mca}
S.~Dulat, T.-J. Hou, J.~Gao, M.~Guzzi, J.~Huston, P.~Nadolsky, J.~Pumplin,
  C.~Schmidt, D.~Stump, and C.~P. Yuan, ``{New parton distribution functions
  from a global analysis of quantum chromodynamics},''
  \href{http://dx.doi.org/10.1103/PhysRevD.93.033006}{{\em Phys. Rev.}
  {\bfseries D93} (2016) 033006},
\href{http://arxiv.org/abs/1506.07443}{{\ttfamily arXiv:1506.07443 [hep-ph]}}.

\bibitem{Eskola:2009uj}
K.~J. Eskola, H.~Paukkunen, and C.~A. Salgado, ``{EPS09: A New Generation of
  NLO and LO Nuclear Parton Distribution Functions},''
  \href{http://dx.doi.org/10.1088/1126-6708/2009/04/065}{{\em JHEP} {\bfseries
  04} (2009) 065},
\href{http://arxiv.org/abs/0902.4154}{{\ttfamily arXiv:0902.4154 [hep-ph]}}.

\bibitem{Kovarik:2015cma}
K.~Kovarik {\em et~al.}, ``{nCTEQ15 - Global analysis of nuclear parton
  distributions with uncertainties in the CTEQ framework},''
  \href{http://dx.doi.org/10.1103/PhysRevD.93.085037}{{\em Phys. Rev.}
  {\bfseries D93} (2016) 085037},
\href{http://arxiv.org/abs/1509.00792}{{\ttfamily arXiv:1509.00792 [hep-ph]}}.

\bibitem{Armesto:2015lrg}
N.~Armesto, H.~Paukkunen, J.~M. Pen\'in, C.~A. Salgado, and P.~Zurita, ``{An
  analysis of the impact of LHC Run I proton--lead data on nuclear parton
  densities},'' \href{http://dx.doi.org/10.1140/epjc/s10052-016-4078-9}{{\em
  Eur. Phys. J.} {\bfseries C76} (2016) 218},
\href{http://arxiv.org/abs/1512.01528}{{\ttfamily arXiv:1512.01528 [hep-ph]}}.

\bibitem{Helenius:2012wd}
I.~Helenius, K.~J. Eskola, H.~Honkanen, and C.~A. Salgado, ``{Impact-Parameter
  Dependent Nuclear Parton Distribution Functions: EPS09s and EKS98s and Their
  Applications in Nuclear Hard Processes},''
  \href{http://dx.doi.org/10.1007/JHEP07(2012)073}{{\em JHEP} {\bfseries 07}
  (2012) 073},
\href{http://arxiv.org/abs/1205.5359}{{\ttfamily arXiv:1205.5359 [hep-ph]}}.

\end{thebibliography}\endgroup

\newpage
\appendix
\section{The ALICE Collaboration}
\label{app:collab}

\begingroup
\small
\begin{flushleft}
S.~Acharya\Irefn{org137}\And 
D.~Adamov\'{a}\Irefn{org94}\And 
J.~Adolfsson\Irefn{org33}\And 
M.M.~Aggarwal\Irefn{org99}\And 
G.~Aglieri Rinella\Irefn{org34}\And 
M.~Agnello\Irefn{org30}\And 
N.~Agrawal\Irefn{org47}\And 
Z.~Ahammed\Irefn{org137}\And 
S.U.~Ahn\Irefn{org78}\And 
S.~Aiola\Irefn{org141}
% ...
\And
\renewcommand\labelenumi{\textsuperscript{\theenumi}~}

\section*{Affiliation notes}
\renewcommand\theenumi{\roman{enumi}}
\begin{Authlist}
\item \Adef{org*}Deceased
\item \Adef{orgI}Dipartimento DET del Politecnico di Torino, Turin, Italy
\item \Adef{orgII}M.V. Lomonosov Moscow State University, D.V. Skobeltsyn Institute of Nuclear, Physics, Moscow, Russia
\item \Adef{orgIII}Department of Applied Physics, Aligarh Muslim University, Aligarh, India
\item \Adef{orgIV}Institute of Theoretical Physics, University of Wroclaw, Poland
\end{Authlist}

\section*{Collaboration Institutes}
\renewcommand\theenumi{\arabic{enumi}~}
\begin{Authlist}
\item \Idef{org1}A.I. Alikhanyan National Science Laboratory (Yerevan Physics Institute) Foundation, Yerevan, Armenia
\item \Idef{org2}Benem\'{e}rita Universidad Aut\'{o}noma de Puebla, Puebla, Mexico
\item \Idef{org3}Bogolyubov Institute for Theoretical Physics, Kiev, Ukraine
\item \Idef{org4}Bose Institute, Department of Physics  and Centre for Astroparticle Physics and Space Science (CAPSS), Kolkata, India
\item \Idef{org5}Budker Institute for Nuclear Physics, Novosibirsk, Russia
\item \Idef{org6}California Polytechnic State University, San Luis Obispo, California, United States
\item \Idef{org7}Central China Normal University, Wuhan, China
\item \Idef{org8}Centre de Calcul de l'IN2P3, Villeurbanne, Lyon, France
\item \Idef{org9}Centro de Aplicaciones Tecnol\'{o}gicas y Desarrollo Nuclear (CEADEN), Havana, Cuba
\item \Idef{org10}Centro de Investigaci\'{o}n y de Estudios Avanzados (CINVESTAV), Mexico City and M\'{e}rida, Mexico
\item \Idef{org11}Centro Fermi - Museo Storico della Fisica e Centro Studi e Ricerche ``Enrico Fermi', Rome, Italy
\item \Idef{org12}Chicago State University, Chicago, Illinois, United States
\item \Idef{org13}China Institute of Atomic Energy, Beijing, China
\item \Idef{org14}Chonbuk National University, Jeonju, Republic of Korea
\item \Idef{org15}COMSATS Institute of Information Technology (CIIT), Islamabad, Pakistan
\item \Idef{org16}Department of Physics, Aligarh Muslim University, Aligarh, India
\item \Idef{org17}Department of Physics, Ohio State University, Columbus, Ohio, United States
\item \Idef{org18}Department of Physics, Pusan National University, Pusan, Republic of Korea
\item \Idef{org19}Department of Physics, Sejong University, Seoul, Republic of Korea
\item \Idef{org20}Department of Physics, University of Oslo, Oslo, Norway
\item \Idef{org21}Department of Physics and Technology, University of Bergen, Bergen, Norway
\item \Idef{org22}Dipartimento di Fisica dell'Universit\`{a} 'La Sapienza' and Sezione INFN, Rome, Italy
\item \Idef{org23}Dipartimento di Fisica dell'Universit\`{a} and Sezione INFN, Cagliari, Italy
\item \Idef{org24}Dipartimento di Fisica dell'Universit\`{a} and Sezione INFN, Trieste, Italy
\item \Idef{org25}Dipartimento di Fisica dell'Universit\`{a} and Sezione INFN, Turin, Italy
\item \Idef{org26}Dipartimento di Fisica e Astronomia dell'Universit\`{a} and Sezione INFN, Bologna, Italy
\item \Idef{org27}Dipartimento di Fisica e Astronomia dell'Universit\`{a} and Sezione INFN, Catania, Italy
\item \Idef{org28}Dipartimento di Fisica e Astronomia dell'Universit\`{a} and Sezione INFN, Padova, Italy
\item \Idef{org29}Dipartimento di Fisica `E.R.~Caianiello' dell'Universit\`{a} and Gruppo Collegato INFN, Salerno, Italy
\item \Idef{org30}Dipartimento DISAT del Politecnico and Sezione INFN, Turin, Italy
\item \Idef{org31}Dipartimento di Scienze e Innovazione Tecnologica dell'Universit\`{a} del Piemonte Orientale and INFN Sezione di Torino, Alessandria, Italy
\item \Idef{org32}Dipartimento Interateneo di Fisica `M.~Merlin' and Sezione INFN, Bari, Italy
\item \Idef{org33}Division of Experimental High Energy Physics, University of Lund, Lund, Sweden
\item \Idef{org34}European Organization for Nuclear Research (CERN), Geneva, Switzerland
\item \Idef{org35}Excellence Cluster Universe, Technische Universit\"{a}t M\"{u}nchen, Munich, Germany
\item \Idef{org36}Faculty of Engineering, Bergen University College, Bergen, Norway
\item \Idef{org37}Faculty of Mathematics, Physics and Informatics, Comenius University, Bratislava, Slovakia
\item \Idef{org38}Faculty of Nuclear Sciences and Physical Engineering, Czech Technical University in Prague, Prague, Czech Republic
\item \Idef{org39}Faculty of Science, P.J.~\v{S}af\'{a}rik University, Ko\v{s}ice, Slovakia
\item \Idef{org40}Faculty of Technology, Buskerud and Vestfold University College, Tonsberg, Norway
\item \Idef{org41}Frankfurt Institute for Advanced Studies, Johann Wolfgang Goethe-Universit\"{a}t Frankfurt, Frankfurt, Germany
\item \Idef{org42}Gangneung-Wonju National University, Gangneung, Republic of Korea
\item \Idef{org43}Gauhati University, Department of Physics, Guwahati, India
\item \Idef{org44}Helmholtz-Institut f\"{u}r Strahlen- und Kernphysik, Rheinische Friedrich-Wilhelms-Universit\"{a}t Bonn, Bonn, Germany
\item \Idef{org45}Helsinki Institute of Physics (HIP), Helsinki, Finland
\item \Idef{org46}Hiroshima University, Hiroshima, Japan
\item \Idef{org47}Indian Institute of Technology Bombay (IIT), Mumbai, India
\item \Idef{org48}Indian Institute of Technology Indore, Indore, India
\item \Idef{org49}Indonesian Institute of Sciences, Jakarta, Indonesia
\item \Idef{org50}INFN, Laboratori Nazionali di Frascati, Frascati, Italy
\item \Idef{org51}INFN, Sezione di Bari, Bari, Italy
\item \Idef{org52}INFN, Sezione di Bologna, Bologna, Italy
\item \Idef{org53}INFN, Sezione di Cagliari, Cagliari, Italy
\item \Idef{org54}INFN, Sezione di Catania, Catania, Italy
\item \Idef{org55}INFN, Sezione di Padova, Padova, Italy
\item \Idef{org56}INFN, Sezione di Roma, Rome, Italy
\item \Idef{org57}INFN, Sezione di Torino, Turin, Italy
\item \Idef{org58}INFN, Sezione di Trieste, Trieste, Italy
\item \Idef{org59}Inha University, Incheon, Republic of Korea
\item \Idef{org60}Institut de Physique Nucl\'eaire d'Orsay (IPNO), Universit\'e Paris-Sud, CNRS-IN2P3, Orsay, France
\item \Idef{org61}Institute for Nuclear Research, Academy of Sciences, Moscow, Russia
\item \Idef{org62}Institute for Subatomic Physics of Utrecht University, Utrecht, Netherlands
\item \Idef{org63}Institute for Theoretical and Experimental Physics, Moscow, Russia
\item \Idef{org64}Institute of Experimental Physics, Slovak Academy of Sciences, Ko\v{s}ice, Slovakia
\item \Idef{org65}Institute of Physics, Academy of Sciences of the Czech Republic, Prague, Czech Republic
\item \Idef{org66}Institute of Physics, Bhubaneswar, India
\item \Idef{org67}Institute of Space Science (ISS), Bucharest, Romania
\item \Idef{org68}Institut f\"{u}r Informatik, Johann Wolfgang Goethe-Universit\"{a}t Frankfurt, Frankfurt, Germany
\item \Idef{org69}Institut f\"{u}r Kernphysik, Johann Wolfgang Goethe-Universit\"{a}t Frankfurt, Frankfurt, Germany
\item \Idef{org70}Institut f\"{u}r Kernphysik, Westf\"{a}lische Wilhelms-Universit\"{a}t M\"{u}nster, M\"{u}nster, Germany
\item \Idef{org71}Instituto de Ciencias Nucleares, Universidad Nacional Aut\'{o}noma de M\'{e}xico, Mexico City, Mexico
\item \Idef{org72}Instituto de F\'{i}sica, Universidade Federal do Rio Grande do Sul (UFRGS), Porto Alegre, Brazil
\item \Idef{org73}Instituto de F\'{\i}sica, Universidad Nacional Aut\'{o}noma de M\'{e}xico, Mexico City, Mexico
\item \Idef{org74}IRFU, CEA, Universit\'{e} Paris-Saclay, Saclay, France
\item \Idef{org75}iThemba LABS, National Research Foundation, Somerset West, South Africa
\item \Idef{org76}Joint Institute for Nuclear Research (JINR), Dubna, Russia
\item \Idef{org77}Konkuk University, Seoul, Republic of Korea
\item \Idef{org78}Korea Institute of Science and Technology Information, Daejeon, Republic of Korea
\item \Idef{org79}KTO Karatay University, Konya, Turkey
\item \Idef{org80}Laboratoire de Physique Subatomique et de Cosmologie, Universit\'{e} Grenoble-Alpes, CNRS-IN2P3, Grenoble, France
\item \Idef{org81}Lawrence Berkeley National Laboratory, Berkeley, California, United States
\item \Idef{org82}Moscow Engineering Physics Institute, Moscow, Russia
\item \Idef{org83}Nagasaki Institute of Applied Science, Nagasaki, Japan
\item \Idef{org84}Nara Women{'}s University (NWU), Nara, Japan
\item \Idef{org85}National and Kapodistrian University of Athens, Physics Department, Athens, Greece
\item \Idef{org86}National Centre for Nuclear Studies, Warsaw, Poland
\item \Idef{org87}National Institute for Physics and Nuclear Engineering, Bucharest, Romania
\item \Idef{org88}National Institute of Science Education and Research, HBNI, Jatni, India
\item \Idef{org89}National Nuclear Research Center, Baku, Azerbaijan
\item \Idef{org90}National Research Centre Kurchatov Institute, Moscow, Russia
\item \Idef{org91}Niels Bohr Institute, University of Copenhagen, Copenhagen, Denmark
\item \Idef{org92}Nikhef, Nationaal instituut voor subatomaire fysica, Amsterdam, Netherlands
\item \Idef{org93}Nuclear Physics Group, STFC Daresbury Laboratory, Daresbury, United Kingdom
\item \Idef{org94}Nuclear Physics Institute, Academy of Sciences of the Czech Republic, \v{R}e\v{z} u Prahy, Czech Republic
\item \Idef{org95}Oak Ridge National Laboratory, Oak Ridge, Tennessee, United States
\item \Idef{org96}Petersburg Nuclear Physics Institute, Gatchina, Russia
\item \Idef{org97}Physics Department, Creighton University, Omaha, Nebraska, United States
\item \Idef{org98}Physics department, Faculty of science, University of Zagreb, Zagreb, Croatia
\item \Idef{org99}Physics Department, Panjab University, Chandigarh, India
\item \Idef{org100}Physics Department, University of Cape Town, Cape Town, South Africa
\item \Idef{org101}Physics Department, University of Jammu, Jammu, India
\item \Idef{org102}Physics Department, University of Rajasthan, Jaipur, India
\item \Idef{org103}Physikalisches Institut, Eberhard Karls Universit\"{a}t T\"{u}bingen, T\"{u}bingen, Germany
\item \Idef{org104}Physikalisches Institut, Ruprecht-Karls-Universit\"{a}t Heidelberg, Heidelberg, Germany
\item \Idef{org105}Physik Department, Technische Universit\"{a}t M\"{u}nchen, Munich, Germany
\item \Idef{org106}Research Division and ExtreMe Matter Institute EMMI, GSI Helmholtzzentrum f\"ur Schwerionenforschung GmbH, Darmstadt, Germany
\item \Idef{org107}Rudjer Bo\v{s}kovi\'{c} Institute, Zagreb, Croatia
\item \Idef{org108}Russian Federal Nuclear Center (VNIIEF), Sarov, Russia
\item \Idef{org109}Saha Institute of Nuclear Physics, Kolkata, India
\item \Idef{org110}School of Physics and Astronomy, University of Birmingham, Birmingham, United Kingdom
\item \Idef{org111}Secci\'{o}n F\'{\i}sica, Departamento de Ciencias, Pontificia Universidad Cat\'{o}lica del Per\'{u}, Lima, Peru
\item \Idef{org112}SSC IHEP of NRC Kurchatov institute, Protvino, Russia
\item \Idef{org113}Stefan Meyer Institut f\"{u}r Subatomare Physik (SMI), Vienna, Austria
\item \Idef{org114}SUBATECH, IMT Atlantique, Universit\'{e} de Nantes, CNRS-IN2P3, Nantes, France
\item \Idef{org115}Suranaree University of Technology, Nakhon Ratchasima, Thailand
\item \Idef{org116}Technical University of Ko\v{s}ice, Ko\v{s}ice, Slovakia
\item \Idef{org117}Technical University of Split FESB, Split, Croatia
\item \Idef{org118}The Henryk Niewodniczanski Institute of Nuclear Physics, Polish Academy of Sciences, Cracow, Poland
\item \Idef{org119}The University of Texas at Austin, Austin, Texas, United States
\item \Idef{org120}Universidad Aut\'{o}noma de Sinaloa, Culiac\'{a}n, Mexico
\item \Idef{org121}Universidade de S\~{a}o Paulo (USP), S\~{a}o Paulo, Brazil
\item \Idef{org122}Universidade Estadual de Campinas (UNICAMP), Campinas, Brazil
\item \Idef{org123}Universidade Federal do ABC, Santo Andre, Brazil
\item \Idef{org124}University of Houston, Houston, Texas, United States
\item \Idef{org125}University of Jyv\"{a}skyl\"{a}, Jyv\"{a}skyl\"{a}, Finland
\item \Idef{org126}University of Liverpool, Liverpool, United Kingdom
\item \Idef{org127}University of Tennessee, Knoxville, Tennessee, United States
\item \Idef{org128}University of the Witwatersrand, Johannesburg, South Africa
\item \Idef{org129}University of Tokyo, Tokyo, Japan
\item \Idef{org130}University of Tsukuba, Tsukuba, Japan
\item \Idef{org131}Universit\'{e} Clermont Auvergne, CNRS/IN2P3, LPC, Clermont-Ferrand, France
\item \Idef{org132}Universit\'{e} de Lyon, Universit\'{e} Lyon 1, CNRS/IN2P3, IPN-Lyon, Villeurbanne, Lyon, France
\item \Idef{org133}Universit\'{e} de Strasbourg, CNRS, IPHC UMR 7178, F-67000 Strasbourg, France, Strasbourg, France
\item \Idef{org134}Universit\`{a} degli Studi di Pavia, Pavia, Italy
\item \Idef{org135}Universit\`{a} di Brescia, Brescia, Italy
\item \Idef{org136}V.~Fock Institute for Physics, St. Petersburg State University, St. Petersburg, Russia
\item \Idef{org137}Variable Energy Cyclotron Centre, Kolkata, India
\item \Idef{org138}Warsaw University of Technology, Warsaw, Poland
\item \Idef{org139}Wayne State University, Detroit, Michigan, United States
\item \Idef{org140}Wigner Research Centre for Physics, Hungarian Academy of Sciences, Budapest, Hungary
\item \Idef{org141}Yale University, New Haven, Connecticut, United States
\item \Idef{org142}Yonsei University, Seoul, Republic of Korea
\item \Idef{org143}Zentrum f\"{u}r Technologietransfer und Telekommunikation (ZTT), Fachhochschule Worms, Worms, Germany
\end{Authlist}
\endgroup
\end{document}